\documentclass[lettersize,journal]{IEEEtran}

\usepackage{cite}
\usepackage[utf8]{inputenc}
\usepackage{amsmath,amssymb,amsfonts}
\usepackage{algorithm}
\usepackage{algpseudocode}
\usepackage{graphicx}
\usepackage{textcomp,soul}
\usepackage{bm}
\usepackage{caption}
\usepackage{subcaption}

\begin{document}

\title{Fractional Doppler Effects on OTFS-NOMA HetNets with Mixed-Mobility Users}

\author{\IEEEauthorblockN{Wafa Hedhly, Leila Musavian, and Nikolaos Thomos}
 \thanks{The authors are with the School of Computer Science and Electronic Engineering, University of Essex, Wivenhoe Park, Colchester CO43SQ, United Kingdom (e-mail: { \{wafa.hedhly, leila.musavian, nthomos\}@essex.ac.uk).} This work was funded by the Engineering and Physical Science Research Council, UK, [grants number EP/X012204/1, EP/X04047X/2, EP/Y037243/1].
 }
 }

\maketitle
\begin{abstract}
Heterogeneous networks (HetNets) are considered a promising approach to meet the increasing throughput requirements of 6G vehicular networks.
The integration of orthogonal time frequency space (OTFS) modulation and non-orthogonal multiple access (NOMA) has demonstrated significant improvements in the reliability of wireless networks under mixed-mobility. Applying this combination in HetNets helps accommodate high-mobility (HM) and low-mobility (LM) users while effectively addressing high Doppler shifts. However, in practical scenarios, fractional Doppler arises from the mismatch between the actual Doppler frequency and its discrete representation on the DD grid. This effect leads to inter-Doppler interference (IDI) that can significantly degrade detection performance.
In our work, we investigate the effect of fractional Doppler on OTFS-NOMA systems under mixed-mobility conditions and analyze how NOMA power allocation can be optimized to mitigate the resulting performance degradation.
OTFS modulation is used for the HM user, while the LM users’ symbols are embedded in the time-frequency (TF) domain. Minimum mean square error (MMSE) detection is utilized under multiple assumptions regarding the knowledge of the IDI parameters. We investigate the impact of optimizing NOMA power allocation on the overall system performance. {The purpose of this work is to analyze the impact of fractional Doppler in OTFS-NOMA systems in HetNets rather than proposing new detection algorithms for OTFS modulation.
The results validate the significant performance impact of fractional Doppler on the system and demonstrate the role of power allocation in mitigating IDI effects in terms of spectral efficiency and number of users served.}
\end{abstract}


\section{Introduction}
{{T}{he} never-ending human attempts to enhance connectivity around the globe have resulted in a complex communication ecosystem, with 6G networks envisioned to integrate a variety of technologies, including vehicular networks, alongside aerial networks, satellites, optic fibers, etc. }
Consequently, in modern networks, a plethora of communication scenarios, challenges, and requirements have emerged, consistently pushing the advances in communication technologies. Nowadays, it is more common to have users with different quality-of-service (QoS) requirements, mobility profiles, and available resources, leading to an unavoidable heterogeneity that cannot be addressed by homogeneous technologies or classical approaches. 
Furthermore, this explosive growth in applications and growing diversity has brought about unprecedented and pressing performance requirements that demand more efficient resource allocation and network adaptability.
In this regard, heterogeneous networks (HetNets) address these increasing demands by balancing macro station loads through smaller, low-power cells alongside existing macrocells \cite{xu2021survey}.
This novel redirection of the traffic helps offload the cellular networks, achieve higher throughput and improve the provided services to particular and emerging urban hotspots. As expected, this collaboration of cells and joint utilization of resources comes at the expense of increased complexity and requires rigorous interference mitigation \cite{lopez2011enhanced}, handover management \cite{tong2021mobility} and
efficient resource allocation techniques \cite{miao2012joint}. 
{In HetNets, users with diversified mobility levels and profiles coexist within the coverage of the same cell and share the same resources, introducing significant challenges in resource allocation, interference mitigation, load balancing and users association between different cells.} 

 Users moving at high speeds face multiple challenges arising from their doubly-dispersive channels that substantially affect their throughput and compromise their performance. These dynamics of the wireless channel result in multiple challenges in channel estimation, signal equalization and detection and increased bit-error rates (BERs).
Under such circumstances, orthogonal frequency division multiplexing (OFDM) suffers from power leakage between adjacent subcarriers, leading to inter-carrier interference (ICI) \cite{singh2022ber}. To cope with these limitations, orthogonal time frequency space (OTFS) has been widely adopted as a next-generation modulation scheme that can circumvent the shortcomings of OFDM by transforming the fast-varying multipath channel to a quasi-stationary two-dimensional representation in the delay-Doppler (DD) domain \cite{singh2022ber}. 
As a result, the interest in OTFS modulation has been rapidly increasing, and there is significant research about its potential in high mobility scenarios for internet-of-things (IoT) \cite{xiao2021overview}, satellite communications \cite{wang2022joint}, space-air-ground integrated networks (SAGIN) \cite{xu2022otfs},
integrated sensing and communications (ISAC) \cite{yuan2021integrated}, etc. 
However, the Doppler taps of the channel representation in the DD domain are not generally integer and fall on a non-integer value between two successive values. This results in fractional Doppler and an inevitable inter-Doppler interference (IDI) \cite{raviteja2018interference}. The conventional implementation of OTFS in the literature does not take this fractional shift into account, leading to detection errors. 
This problem can be partly solved by increasing the OTFS resolution which increases the likelihood of the channel taps falling on an integer Doppler tap. However, it is known that one of the disadvantages of OTFS modulation is the increased latency compared to that of OFDM. 
Hence, implementing OTFS modulation can bring a performance boost in high-mobility networks as long as the IDI is properly assessed to propose suitable coping mechanisms.

Following the same rationale, OTFS modulation performance depends on the frame dimension since it spans the whole resource block. Therefore, using orthogonal access schemes requires sacrificing either temporal or frequency resources. On this account, a combination of OTFS and non-orthogonal multiple access (NOMA) has been considered a plausible approach that allows the OTFS frame to span the whole available resource block without compromising the system's spectral efficiency \cite{saito2013non}. Moreover, an OTFS-NOMA system reaps the advantages of non-orthogonal access in increasing the spectral efficiency of the system by accommodating multiple users in the same resource block if efficient successive interference cancellation is implemented. In such systems, OTFS modulation provides Doppler resilience to high-mobility (HM) users, and NOMA helps accommodate stationary and low-mobility (LM) users with HM users efficiently. 
Furthermore, the NOMA power allocation factors represent an additional flexible set of parameters that can be strategically set to adapt the system to users' needs.
These merits promote exploiting OTFS-NOMA with its versatile and flexible attributes to HetNets, where a variety of throughput demands and user profiles coexist in the same geographical~area.

{Different from previous works about OTFS-NOMA \cite{ding2019otfs,ding2019robust,ge2021otfs,zhou2022active,zhou2021joint,hu2024cross,mcwade2023low,hedhly2024otfs,hedhly2024performance,li2025pattern,zhao2025integrated,zhou2025resource}, in this paper, we implement OTFS-NOMA in a HetNet architecture where mobile users (MUs) are HM or LM users and while taking into account the impact of fractional Doppler on the system performance. We analyze and investigate the impact of IDI on the system performance. We also propose a resource allocation framework to improve the system operation. 
The MUs are within the coverage of a cellular macro base station (MBS) where multiple pico base stations (PBSs) provide access links to the MUs.} We follow the NOMA principle, and hence, the MU receiver (both HM and LM users) detects the HM user's signal. After detecting and subtracting the HM user's signal in the DD domain, the LM user detects its signal in the TF domain. We summarize the contributions as follows:
\begin{itemize}
    \item {We propose an OTFS-NOMA framework for HetNets with mixed-mobility users, where HM and LM users coexist within the same picocell and share radio resources through NOMA. Resource allocation strategies in terms of bandwidth partitioning and user assignment to PBSs are provided. The implementation of OTFS-NOMA in HetNets has not been investigated in previous research works.}
    \item {We investigate, to the best of our knowledge for the first time, the impact of fractional Doppler and IDI on OTFS-NOMA HetNets and quantify the resulting performance degradation under mixed-mobility conditions. We consider the fractional Doppler characteristics in the received signal when modeling the proposed OTFS-NOMA system resulting in an IDI signal and the IDI-free signal. This paper focuses on fractional Doppler rather than fractional delay as it is the main impairment in high-mobility scenarios.}
    \item {We propose two detection approaches to analyze the effect of IDI depending on the availability of channel information and fractional Doppler parameters using minimum mean square error (MMSE) detection. First, we apply the detection matrix without knowledge of the interference matrix parameters resulting from the fractional Doppler. Second, we apply the full detection matrix, where the IDI parameters are assumed to be known at the receiver.} 
    \item We derive the signal-to-noise ratios (SNRs) closed-form expressions considering both detection approaches at the HM side. Moreover, we derive the LM users sum-rate expression after subtracting the HM user's signal.
    \item {We optimize the NOMA power allocation factors with the objective of maximizing the HM users SNR while keeping a minimum sum-rate requirement for the LM users, while implementing Lasso regularization to control the distribution of resources among LM users while satisfying the system's requirements.}
    \item Finally, we provide suitable simulations to evaluate the performance of the proposed OTFS-NOMA system for HetNets. We evaluate the effect of the IDI and the NOMA power allocation factors on the performance of MUs within the network. Furthermore, we evaluate the statistical properties of the system in the presence of the IDI.
    We evaluate the probability density function (PDF) of the detection SNR and the average number of LM users served by the network. We observe that the fractional Doppler significantly affects the spectral efficiency of the HM user, and this depends on the NOMA allocation factor. Moreover, our optimization helps improve user fairness in the network. 
\end{itemize}
It is important to emphasize that this work does not aim to propose new detection algorithms for OTFS or improve OTFS modulation. The main objective is to understand and quantify the performance degradation caused by fractional Doppler in OTFS-NOMA systems and to provide insights into power allocation detection trade-offs in mixed-mobility environments. To the best of our knowledge, the impact of fractional Doppler on OTFS-NOMA remains largely unexplored. {The proposed framework is particularly relevant to high-mobility heterogeneous networks where users with different mobility profiles coexist within the same coverage area. For instance, highways, high-speed railway, and future SAGIN represent suitable deployment scenarios, where high-mobility users experience significant Doppler effects.}
The rest of the paper is organized as follows. Section II presents the system description and system model of the proposed HetNet with OTFS-NOMA. In Section III, we provide the detailed detection method at both HM and LM users. Section IV presents the optimization problem formulation and solution of the NOMA allocation factors. Then, in Section V, we evaluate the system through multiple simulation examples. Finally, the conclusion is provided in Section VI.

\section{{Related Works}}
Recently, there has been a significant increase in research attention towards OTFS-NOMA systems \cite{xiao2021overview,ding2019otfs,ding2019robust,ge2021otfs,zhou2022active,zhou2021joint,hu2024cross,zhao2025integrated,zhou2025resource,li2025pattern,mcwade2023low,hedhly2024otfs}. Ding et al. introduced in their work \cite{ding2019otfs} a system combining OTFS modulation and NOMA to optimize communication for one HM user and multiple LM users. In \cite{ding2019robust}, a robust MIMO beamforming system was proposed while implementing a comparable OTFS-NOMA to the setup described in \cite{ding2019otfs}. 
{Power-domain OTFS-NOMA has been proposed in \cite{xiao2021overview} to explore the potential of OTFS in IoT networks, where OTFS-NOMA outperformed OFDM-NOMA in terms of spectral efficiency.}
In \cite{ge2021otfs}, uplink transmission was examined with the utilization of OTFS modulation for all users combined with NOMA. Zhou et al. utilized OTFS-NOMA in their studies \cite{zhou2022active} and \cite{zhou2021joint} to enable connectivity for a large quantity of IoT devices and reduce the substantial Doppler spread in satellite-to-ground communication. In \cite{hu2024cross}, cross-domain channel estimation was employed to introduce a new method for successive interference cancellation in OTFS-NOMA systems.
In \cite{zhao2025integrated}, the reliability and capacity of an OTFS-NOMA framework was investigated for multi-beam LEO satellite systems. In \cite{zhou2025resource}, the authors studied resource allocation for a mmWave OTFS-NOMA system, where power and beamwidth are optimized to improve throughput. The authors of \cite{li2025pattern} proposed a pattern-domain NOMA scheme for OTFS modulation with a low-complexity receiver design to enhance spectral efficiency.
A low-complexity method for user equalization and detection in OTFS-NOMA systems, taking into account fractional Doppler, was introduced by McWade et al. in their study \cite{mcwade2023low}. 
The work in \cite{hedhly2024otfs} discussed an OTFS-NOMA setup in which a base station employs MIMO beamforming to communicate with HM and LM users grouped into clusters. 
{Although several works consider non-zero fractional Dopplers, they do not explicitly investigate the impact of fractional Doppler, IDI modeling and its interplay with NOMA parameters which are the key focus of our study.}
{In our previous work \cite{hedhly2024performance}, we provided preliminary results on the performance difference between the presence and absence of fractional Doppler in a simplified single-cell scenario of OTFS-NOMA and considering a single HM user. }\\
{Although OTFS-NOMA has been previously investigated in the literature, several important aspects remain largely unexplored. In fact, most existing studies focus on conventional OTFS-NOMA systems without explicitly quantifying the impact of fractional Doppler and the resulting IDI on the system performance. Moreover, the performance of OTFS-NOMA in a HetNet architecture and the interaction between fractional Doppler, NOMA power allocation in the presence of mixed-mobility users have not been investigated.}

\section{System Model}
In this section, we present the general framework of the considered system. We provide a detailed description of the system operation, including multiple access, resource allocation, transmitted signals, wireless channel and received signals.
  \begin{figure}[!h]
	\centering 
	\captionsetup{justification=centering,margin=1cm}
	\includegraphics[width=3.5in]{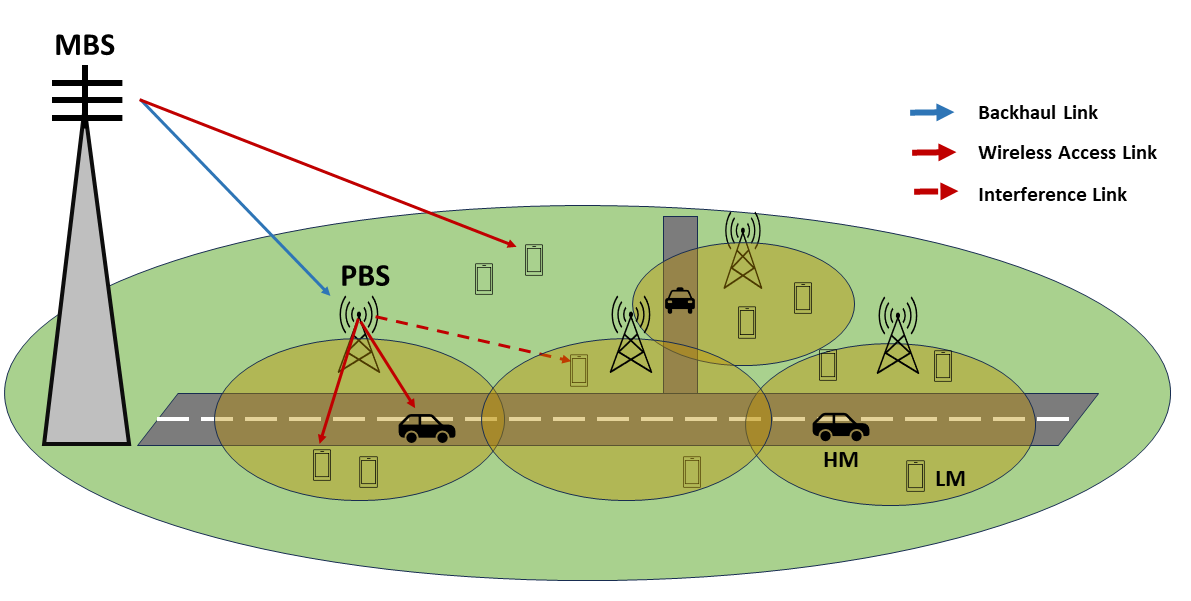}
	\caption{Proposed system setup.}
	\label{system}
\end{figure}
\subsection{System Description}
We consider a network where an MBS equipped with $A_{\mathrm{P}}$ antennas provides wireless coverage to a macrocell. The MBS is connected to the core network through a fiber backhaul. The macrocell is assumed to be dense enough with mobile users to install $Q$ PBS to provide picocells with wireless coverage. MUs are distributed in the network, where users served by the MBS are called macro users and users served by a PBS are called pico users. Each PBS is equipped with $A_{\mathrm{P}}$ antennas while the MUs are equipped with single antennas.
The MBS provides coordination and control services between PBSs through a fiber backhaul link and delivers data packets to MUs through wireless access links. We assume that inter-picocell interference is mitigated at the MBS level. 
The considered HetNet architecture is depicted in Fig. \ref{system}.

Users are distinguished based on their mobility profiles into HM users and LM users.
We assume that $U_q \ge 1$ LM users and a single HM user coexist within the coverage of PBS $q$ where $1\le q \le Q$. We further assume that each HM user is under the coverage of at least one PBS at every instant\footnote{The HM user moves in a specific straight direction like a road or railways}.
LM users access disjoint subcarriers. When a HM user enters the coverage of a picocell, NOMA is deployed to superimpose the signals of the HM user and the existing LM users.
The information symbols of the HM user are represented in the DD domain to form an OTFS frame that spans the whole time-frequency resource block. The information symbols of the LM users are represented in the TF domain. Thus, if the number of time slots is $N$ and the number of subchannels is $M$, then the number of Doppler bins and delay bins of the OTFS frame are $N$ and $M$, respectively.
We assume that a maximum of $M$ LM users can be served by each PBS, which is the number of the available subchannels.
It is worth noting that the paper considers multiple HM and LM users. The assumption of a single HM user per pico cell is representative of practical deployments where a fast-moving user like a vehicle or a train can co-exist with multiple LM users like pedestrians or static devices. The assumption of a single HM user has already been adopted in prior OTFS-NOMA studies \cite{ding2019otfs}. {This assumption also enables us to isolate the impact of fractional Doppler and IDI on OTFS-NOMA performance without introducing additional interference interactions among multiple HM users.}

The total bandwidth $W$ is partitioned into two parts: $\alpha W$ is allocated for the access links of MUs not within the coverage of any PBS, while $(1-\alpha)W$ is allocated for the wireless access communication between each PBS and its MUs. We adopt this bandwidth partitioning to avoid strong interference links from the MBS on the picocell users.
The MBS implements transmit beamforming over $\alpha W$ to communicate with the MUs. Neighboring picocells access different bands (frequency reuse) to limit the interference on the MU from adjacent PBSs.
The HM user's signal and LM users' signals are multiplexed in the same resource block using NOMA. We denote by $\beta_{0,q}$ the power allocation factor of the HM user in picocell $q$. If $p_q$ is the transmitted power in picocell $q$, then, the power allocated for $u_{0,q}$ (HM user in picocell $q$) is $\beta_{0,q} p_q$ while $(1-\beta_{0,q}) p_q$ is distributed between the LM users according to their channel condition. An LM user with a better channel condition gets lower transmit power. 

Each LM user is assigned to a PBS chosen from the pool of $N_q(u)$ candidate PBSs. A candidate PBS for user $u$ is a PBS that satisfies two conditions:
\begin{itemize}
    \item The received signal strength indicator (RSSI) at the user is higher than a minimum threshold to ensure successful demodulation and decoding.
    \item The PBS has at least one free subchannel.
\end{itemize} 
The user measures the RSS of nearby PBSs and shares this information with the MBS. The MBS identifies $N_q(u)$ PBSs as candidates for user $u$ following the rules below.
\begin{itemize}
    \item  If $N_q(u) \ge 1$, the user connects to the PBS with the strongest RSS.
    \item If $N_q(u) = 0$ {(no PBS can be assigned to the user)}, the user is assigned to an access link from the MBS over the dedicated separate band. 
\end{itemize}
In this work, we assume the area is not densely populated by LM users. Thus, the probability of occurrences of the event $N_q(u) = 0$ is considered to be negligible, i.e., all PBS have free subchannels at all times. Therefore, if $U_{\max}$ denotes the maximum number of LM users per picocell, then we assume $U_{\max} << M$. {Since each LM user occupies a dedicated subchannel, this assumption guarantees sufficient frequency resources, allowing the impact of fractional Doppler to be isolated from limitations related to insufficient resources.}

\subsection{Transmitted Signals}
In this work, we are interested in studying the communication links inside the picocells.
We consider $NM$ information symbols transmitted to the HM user are represented in the DD domain. Each packet has a bandwidth of $M \Delta f$ and a duration of $NT$, where $\Delta f$ is the subcarrier spacing and $T = 1/\Delta f$ is the sampling time. 
The DD domain signal $ s_{0,q}^{\mathrm{DD}}$ of the HM user in picocell $q$ is converted to the TF domain as follows for $1 \le n \le N$ and $1 \le m \le M$,
\begin{equation}
    s_{0,q}^{\mathrm{TF}}[n,m] = \frac{1}{NM} \sum_{k = 0}^{N-1} \sum_{l = 0}^{M-1} s_{0,q}^{\mathrm{DD}}[k,l] e^{j2\pi \left( \frac{kn}{N}-\frac{ml}{M}  \right)},
\end{equation}
where $k$ is the Doppler bin and $l$ is the delay bin. 
The base-band TF signal of LM user $u$ of picocell $q$ in time slot $n$ and subchannel $m$ is expressed as in \cite{ding2019otfs},
\begin{align}\label{suq}
    \mathbf{s}_{u,q}^{\mathrm{TF}} [n,m] = 
    \begin{cases}
        & \mathbf{s}_{u,q}^\mathrm{TF}[n, u-1], \quad \mathrm{if} \; m = u - 1, \\
        & 0, \quad \mathrm{otherwise}.
    \end{cases}
\end{align}
In the following, we remove the $\mathrm{DD}$ superscript for the ease of notations.
 The users in each picocell access the spectrum using NOMA paradigm. As a result, the base-band signal of PBS $q$ is expressed in the DD domain as, 
\begin{equation}
    {s}_q[k,l] = \sum_{u = 0}^{U_q} \sqrt{\beta_{u,q}} {s}_{u,q}[k,l],
\end{equation}
where ${s}_{u,q}$ and $\beta_{u,q}$ are the information-bearing signal for user $u$ in picocell $q$ and its corresponding power allocation factor, respectively.
The signal transmitted at picocell $q$ is~expressed~as,
\begin{equation}\label{xa}
    \mathbf{x}_q[k,l] =  \mathbf{v}_{q} {s}_{q}[k,l],
\end{equation}
where $\mathbf{v}_{q} = [v_{1,q}, \ldots ,v_{A,q}]^T$, $v_{a,q}$ is a weighing factor for antenna element $a$.

\subsection{Wireless Channel}
We assume $P_{u,q}$ the number of paths of the wireless channel of user $u$ of picocell $q$, $0 \le u \le U_q$. {The LM users are not subject to Doppler shifts since users moving at low speeds experience negligible Doppler shifts compared to HM users.} The path delays of LM user $u$ are denoted by $\tau_{p,u,q}$ that depends on the transmission distance. 

When fractional Doppler is taken into account, each scatterer in the medium results in multiple subpaths, including one dominant path.
We assume that each path $j$ of the HM user channel has $\Gamma_{j,q}$ subpaths. 
The Doppler and delay taps of path $1 \le j \le P_{0,q}$ of the HM user channel of picocell $q$ are expressed as,
\begin{equation}\label{taps}
    \nu_{s_j,0,q} = \frac{k_{s_j,0,q} + \kappa_{s_j,0,q}}{NT} \quad \mathrm{and} \quad \tau_{j,0,q} = \frac{l_{j,0,q}}{M \Delta f},
\end{equation}
where $l_{j,0,q}$ are the delay tap of path $j$, $k_{s_j,0,q}$  is the Doppler tap of subpath $s_j$ originating from path $j$ and $\kappa_{s_j,0,q}$ is a real number in the interval $	( -1/2,1/2 ] \,$ which represents the fractional Doppler of subpath $s_j$. The parameter $\kappa_{s_j,0,q}$ describes the fractional shift from the nearest Doppler bin. 
{While fractional delay could introduce some minor distortions, its impact is generally less critical than that of fractional Doppler in most wideband systems} \cite{raviteja2018interference}. {Fractional Doppler is usually the dominant impairment in high-mobility scenarios because it directly introduces IDI.}

The corresponding channel impulse responses of the HM user and LM users in picocell $q$ are expressed as follows,
\begin{align}
    h_{0,q} (\tau,\nu) &= \sum_{j = 1}^{P_{0,q}} \sum_{s_j = 1}^{\Gamma_{j,q}} \alpha_{s_j,0,q} \delta \left(\tau - \tau_{j,0,q}  \right) \delta \left( \nu - \nu_{s_j,0,q} \right), \nonumber \\
    h_{u,q}(\tau) &= \sum_{j = 1}^{P_{u,q}} \alpha_{j,u,q} \delta \left( \tau - \tau_{j,u,q} \right), \quad 1 \le u \le U_q,
\end{align}
where $\alpha_{s_j,0,q}$ is the complex channel gain of subpath $s_j$ of the HM user \cite{shi2021deterministic} and $\alpha_{j,u,q}$ is the complex channel gain of path $j$ of LM user $u$.\\
Similarly to \cite{raviteja2018interference} and \cite{shi2021deterministic}, we assume that the Doppler taps $k_{s_j,0,q}$ and fractional shifts $\kappa_{s_j,0,q}$ to be equal for all subpaths generating from the same path, i.e., $\kappa_{s_j,0,q} = \kappa_{j,0,q}$ and $k_{s_j,0,q} = k_{j,0,q}$ for all $j$. This can be justified by the fact that all subpaths corresponding to the same path are generated from the same reflector or cluster of reflectors, leading to negligible differences in terms of angle-of-arrivals. \\

\textbf{Remarks}
\begin{itemize}
    \item When fractional Doppler is not considered, Doppler frequencies are assumed to ideally fall on an exact Doppler bin. Such assumptions can only be valid with very high OTFS grid resolutions, leading to high OTFS frame dimensions and increased processing complexity.
    \item Increasing the signal duration $NT$ helps reduce the fractional Doppler, but it leads to increased OTFS latency \cite{wei2022orthogonal}.
    \item The number of subpaths is significantly smaller than $NM$ which preserves the sparsity characteristic of the OTFS channel matrix.
    \item {We consider fixed delay and Doppler tap locations within each channel realization because the dominant reflectors are assumed unchanged during the observation interval. Consequently, only the path gains vary from one realization to another while the tap locations remain fixed.} Therefore, in each channel realization, the channel OTFS matrix changes only in element gains but not positions.
    \item If for all paths $j$, $\Gamma_{j,q} = 1$, meaning that we do not take into account IDI, the performance at the receiver decreases because of detection errors. 
    \item In the presence of fractional Doppler, in order to acquire full channel state information (CSI), there are $2 P_{0,q}$ additional parameters to estimate at each PBS $q$: fractional shifts $\kappa_{j,0,q} $ and $\Gamma_{j,q}$ for all $P_{0,q}$ paths of the HM user. Without the assumption of similar fractional shifts for all subpaths of the same path, the number of estimated parameters becomes $P_{0,q} + \sum_{j = 1}^{P_{0,q}} \Gamma_{j,q}$.
\end{itemize}

\subsection{Received Signals}
\subsubsection{Received Signal at HM user}
Taking into consideration the fractional Doppler in the channel, the received signal at the HM user of picocell $q$ from antenna element $a$, $1 \le a \le A$ is expressed in the DD domain \cite{raviteja2018interference} as follows,
\begin{align}\label{y0a}
    {y}_{a,0,q}[k,l] &= \sum _{j = 1}^{P_{0,q}} \sum _{s = -N_{j,q}}^{N_{j,q}}  h_{j,a,0,q} (s) \times \,  \nonumber \\& {x}_{a,q} \left [{\left(k - k_{j,0,q} + s \right)_{N}, \left(l - l_{j,0,q}\right)_{M}}\right]\!+ w_{0,q}[k,l],
\end{align}
where $w_{0,q}$ is the complex Gaussian additive noise with zero-mean and variance $\sigma_{\mathrm{w}}^2$, the number $N_{j,q}$ indicates the number of interfering signals defined such that $\Gamma_{j,q} = 2 N_{j,q} + 1 $ and $h_{j,a,0,q}$ is the channel gain expressed as,
\begin{align}\label{hp0}
    h_{j,a,0,q}(s) &= \alpha_{j,a,0,q} \; e^{-j2\pi \nu_{j,0,q} \tau_{j,0,q} } \; \times \nonumber \\
    &{\frac {e^{-j {2\pi } (-s - \kappa _{j,0,q}) }-1}{N e^{-j \frac {2\pi }{N} (- s - \kappa _{j,0,q})}-N}},
\end{align} 
where $\alpha_{j,a,0,q} \sim \mathcal{CN} (0,1/P_{0,q})$ is the channel gain between antenna $a$ of the PBS and the HM user. Herein, we assume that the delay and Doppler taps locations are the same for all antennas. {Since the PBS antenna elements are colocated, the propagation paths observed by different antennas experience nearly identical delay and Doppler locations.}

The number $\Gamma_{j,q}$ describes the number of transmitted signals in the $j$-th path, where $N_{j,q} << N$. We notice that the channel gain $h_{j,a,0,q}(s)$ is a decreasing function of $|s|$. Thus, the signal indexed by $s = 0$ is the signal not affected by fractional Doppler and has the major contribution to the received signal, and corresponds to the dominant subpath. The remaining $2 N_{j,q}$ signals result in interference on neighboring symbols (Doppler taps) in the OTFS grid leading to IDI.

In order to evaluate the IDI, the received signal \eqref{y0a} can be distinguished into two parts: {IDI-free signal and IDI signal} as follows,
\begin{equation}
   {y}_{0,a,q} = {y}_{\mathrm{M},a,q} + y_{\mathrm{I},a,q} + w_{0,q},
\end{equation}
where the two signals ${y}_{\mathrm{M},a,q}$ and $y_{\mathrm{I},a,q}$ are given by,
\begin{align}\label{M_IDI}
    &{y}_{\mathrm{M},a,q}[k,l] = \sum _{p = 1}^{P_{0,q}}  h_{p,a,0,q} (0) {x}_{a,q} \left [{\left(k - k_{p,0,q} \right)_{N} \!, \! \left(l - l_{p,0,q} \right)_{M}}\right] \nonumber \\
    &{y}_{\mathrm{I},a,q}[k,l] = \sum _{p = 1}^{P_{0,q}} \sum _{  \substack{s = -N_{p,q} \\ s \neq 0}}^{N_{p,q}}  h_{p,a,0,q} (s)  \times \nonumber \\  & \qquad \qquad \quad  \qquad \qquad {x}_{a,q} \! \left [{ \left(k-k_{p,0,q}+s\right)_{N} \!, \!\left(l - l_{p,0,q} \right)_{M}}\! \right].
\end{align}
In the following, we assume that the number of subpaths $\Gamma_{p,q}$ is the same for all paths of the HM user channel. Thus, $N_{p,q} = N_0$, for all $1 \le p \le P_{0,q}$ and all $1 \le q \le Q$.

Therefore, the received signal at the HM user of picocell $q$ is expressed as,
\begin{equation}
    y_{0,q} = \sum_{a = 1}^A y_{a,0,q}.
\end{equation}

The received signal can be expressed in a vector form as follows,
\begin{equation}\label{y0}
    \mathbf{y}_{0,q} = \sum_{a = 1}^A \mathbf{H}_{\mathrm{M},a,q} \mathbf{x}_{a,q} + \sum_{a = 1}^A \mathbf{H}_{\mathrm{I},a,q} \mathbf{x}_{a,q}  + \mathbf{w}_{0,q},
\end{equation}
where $\mathbf{H}_{\mathrm{M},a,q}, \mathbf{H}_{\mathrm{I},a,q}  \in \mathcal{C}^{N\!M \times N\!M}$ are the main (dominant) and interference block circulant channel matrices between user $u$ and antenna $a$, respectively, constructed from \eqref{M_IDI}. These matrices are $M$ circulant blocks of $N \times N$ circulant matrices. The noise vector in the DD domain is $\mathbf{w}_{0,q} \sim \mathcal{CN}(\mathbf{0}, {\sigma}_{\mathrm{w}}^2 \mathbf{I}_{N \! M})$.\\ 
While constructing the $N \! M \times 1$ signal vector $\mathbf{x}_{a}$ from \eqref{xa}, the following condition is satisfied: the $(k+Nl)$-th element of $\mathbf{x}_{a,q}$ is equal to $x_{a,q} [k,l]$. The vectors $\mathbf{y}_{0,q}$ and $\mathbf{w}_{0,q}$ can be constructed using the same approach.

The received signal in \eqref{y0} can be rewritten equivalently as follows,
\begin{align}\label{y0q}
    \mathbf{y}_{0,q} &= \underbrace{\sqrt{\beta_{0,q}} \sum_{a = 1}^A v_{a,q} \mathbf{H}_{\mathrm{M},a,q} \mathbf{s}_{0,q}}_{\text{{IDI-free signal}}} + \underbrace{\sqrt{\beta_{0,q}} \sum_{a = 1}^A v_{a,q} \mathbf{H}_{\mathrm{I},a,q} \mathbf{s}_{0,q}}_{\text{IDI signals}}  \nonumber \\ 
    &+ \underbrace{\sum_{a = 1}^A v_{a,q} \mathbf{H}_{a,0,q}
 \sum_{i = 1}^{U_q} \sqrt{ \beta_{i,q}} \mathbf{s}_{i,q}}_{\text{LM users signals}}  + \mathbf{w}_{0,q},
\end{align}
where $\mathbf{H}_{a,0,q} = \mathbf{H}_{\mathrm{M},a,q} + \mathbf{H}_{\mathrm{I},a,q} $ is the overall channel matrix constructed from \eqref{y0a}.

Due to the fractional Doppler, the resulting OTFS matrix has fewer non-zero elements than in the case of the absence of the fractional Doppler. This happens as the number of non-zero elements per row and column of the channel matrix is calculated as $N_{\mathrm{z}} = \sum_{p = 1}^{P_{0,q}} (2N_{p,q} + 1)$. 

\subsubsection{Received Signal at LM user}
The received signal at the LM user is not subject to IDI because of the absence of Doppler shift. Therefore, the received signal at user $u, \; 1 \le u \le U_q$ can be expressed as,
\begin{align}\label{yu}
    \mathbf{y}_{u,q}  &= \sqrt{\beta_{0,q}} \sum_{a = 1}^A v_{a,q} \mathbf{H}_{u,a,q} \mathbf{s}_{0,q} \nonumber \\ &+ \sum_{a = 1}^A v_{a,q} \mathbf{H}_{a,u,q}  
 \sum_{i = 1}^{U_q} \sqrt{\beta_{i,q}} \mathbf{s}_{i,q}  + \mathbf{w}_{u,q},
\end{align}
where $\mathbf{w}_{u,q} \sim \mathcal{CN}(\mathbf{0}, {\sigma}_{\mathrm{w}}^2 \mathbf{I}_{N \! M})$ is the noise vector and $\mathbf{H}_{a,u,q}$ is the channel matrix expressed in the DD domain. 
Since LM users do not undergo Doppler shifts, the OTFS representation of their channel matrices simplifies to a block-diagonal matrix.

\section{Users Detection in the Presence of Fractional Doppler}
In this section, we present the detection approach at the receivers of the MUs in picocell $q$. 
As mentioned earlier, fractional Doppler introduces a number of additional channel parameters: (a) the fractional Doppler $\kappa_{p,0,q}$ and (b) the number of subpaths $N_{p,q}$, for all $1 \le p \le P_{0,q}.$ Signal detection depends on the availability of this channel information at the HM receiver. It is worth noting that NOMA systems rely on SIC which significantly increases the complexity of the receiver.
Thus, we adopt a low-complexity linear MMSE detector as a practical system choice to balance performance and complexity.
We consider the following matrix expressions\footnote{Hereafterwards, we drop the $q$ subscript from the notations for better readability.},
\begin{align}\label{H0}
    \mathbf{H}_{\mathrm{M}} &= \sum_{a = 1}^A v_{a} \mathbf{H}_{\mathrm{M},a}, \nonumber \\
    \mathbf{H}_{u} &= \sum_{a = 1}^A v_{a} \mathbf{H}_{a,u}, \quad 0 \le u \le U.
\end{align}

\subsection{Effect of Inter-Doppler Interference on HM Detection}
In this section, we assume the receiver is not aware of the existence of fractional Doppler and does not have information about the fractional Doppler parameters $\kappa_p$ and $N_p$. Thus, we evaluate the effect of IDI on the HM user's detection.
Using the diagonalization of block-circulant matrices as in \cite{singh2022low}, the channel matrix $\mathbf{H}_{\mathrm{M},a}$ is given by,
\begin{equation}
    \mathbf{H}_{\mathrm{M},a} =  \mathbf{\Psi}^\mathrm{H} \mathbf{D}_{\mathrm{M},a} \mathbf{\Psi},
\end{equation}
where $\mathbf{\Psi} = \mathbf{F}_N \otimes \mathbf{F}_M \in \mathbb{C}^{NM \times NM}$ is a unitary matrix, $\mathbf{F}_N$ and $\mathbf{F}_M$ are $N \times N$ and $M \times M$ discrete Fourier transform (DFT) matrices, respectively and $\mathbf{D}_{\mathrm{M},a}$ is a diagonal matrix with elements $\lambda_{a,i}, 1 \le i \le NM$. 

Therefore, since $\mathbf{\Psi} \mathbf{\Psi}^\mathrm{H} = \mathbf{\Psi}^\mathrm{H}\mathbf{\Psi}  = \mathbf{I}_{MN}$, the equivalent matrix in \eqref{H0} can be rewritten as follows,
\begin{equation}
    \mathbf{H}_\mathrm{M} = \mathbf{\Psi}^\mathrm{H} \mathbf{D}_\mathrm{M} \mathbf{\Psi},
\end{equation}
where the diagonal matrix $\mathbf{D}_\mathrm{M}$ is expressed as,
\begin{equation}
    \mathbf{D}_\mathrm{M} =  \sum_{a = 1}^A v_a \mathbf{D}_{\mathrm{M},a}.
\end{equation}
Therefore, the equalization matrix to detect $\mathbf{s}_{0}$ can be written as follows,
\begin{equation}
    \mathbf{G}_0 =  \left( {\mathbf{H}}_\mathrm{M}^\mathrm{H} \mathbf{H}_\mathrm{M}  + \rho \mathbf{I}_{NM}\right)^{-1} {\mathbf{H}}_\mathrm{M}^\mathrm{H},
\end{equation}
where $\rho > 0$.
The equalization matrix $\mathbf{G}_0$ can be re-expressed as,
\begin{equation}
    \mathbf{G}_0 = \mathbf{\Psi}^\mathrm{H}\mathbf{\Delta}_0 \mathbf{\Psi}.
\end{equation}
where the diagonal matrix $\mathbf{\Delta}_0$ can be expressed as,
\begin{equation}
    \mathbf{\Delta}_0 =  \left( {\mathbf{D}}_\mathrm{M}^{\mathrm{H}} \mathbf{D}_\mathrm{M} + \rho \mathbf{I}_{NM} \right)^{-1} {\mathbf{D}}_\mathrm{M}^{\mathrm{H}}.
\end{equation}
Therefore, $\mathbf{\Delta}_0$ has elements $\delta_{0,i}, 1 \le i \le NM$ and are given by,
\begin{equation}\label{delta0}
    \delta_{0,i} = \frac{\sum_{a = 1}^A   \left({v_a} \lambda_{a,i} \right)^*}{ \left |\sum_{a = 1}^A  v_{a} \lambda_{a,i} \right|^2 + \rho }, 1 \le i \le NM.
\end{equation}
Similarly, the matrices $\mathbf{H}_{\mathrm{I},a}$ and $\mathbf{H}_{a,0}$ can be decomposed as follows,
\begin{align}
    \mathbf{H}_{\mathrm{I},a} &= \mathbf{\Psi}^\mathrm{H} \mathbf{D}_{\mathrm{I},a} \mathbf{\Psi}, \nonumber \\
    \mathbf{H}_{a,0} &= \mathbf{\Psi}^\mathrm{H} \mathbf{D}_{a,0} \mathbf{\Psi},
\end{align}
where $\mathbf{D}_{\mathrm{I},a}$ and $\mathbf{D}_{a,0}$ are diagonal matrices comprising the eigenvalues of $\mathbf{H}_{\mathrm{I},a}$ and $\mathbf{H}_{a,0}$, respectively. As a result, the proposed equalizer is applied at the receiver of the HM user in \eqref{y0q} as shown below,
\begin{equation}\label{Gy}
    \mathbf{G}_0 \mathbf{y}_0 = \sqrt{\beta_0} \mathbf{E}_0 \mathbf{s}_0 +\sqrt{\beta_0} \mathbf{F}_0 \mathbf{s}_0  + \mathbf{T}_0 \sum_{i = 1}^{U} \sqrt{\beta_i} \mathbf{s}_i +  \mathbf{z}_0,
\end{equation}
where $\mathbf{z}_0 = \mathbf{G}_0 \mathbf{w}_0 $ and the matrices $\mathbf{F}_0$ and $\mathbf{T}_0$ are given by,
\begin{align}
    \mathbf{E}_0 &= \mathbf{\Psi}^{\mathrm{H}} \mathbf{\Delta_E} \mathbf{\Psi}, \quad \mathbf{\Delta_E} = \mathbf{\Delta}_0 \mathbf{D}_\mathrm{M} \nonumber \\
    \mathbf{F}_0 & = \mathbf{\Psi}^{\mathrm{H}} \mathbf{\Delta_F} \mathbf{\Psi}, \quad \mathbf{\Delta_F} = \mathbf{\Delta}_0 \sum_{a = 1}^A v_a \mathbf{D}_{\mathrm{I},a} \nonumber \\
    \mathbf{T}_0 &= \mathbf{\Psi}^{\mathrm{H}} \mathbf{\Delta_T} \mathbf{\Psi}, \quad \mathbf{\Delta_T} =\mathbf{\Delta}_0 \sum_{a = 1}^A v_a  \mathbf{D}_{a,0}.
\end{align}

Equivalently, the received signal in \eqref{Gy} can be re-expressed as follows,
\begin{equation}\label{Gy1}
     \mathbf{G}_0 \mathbf{y}_0 = \sqrt{\beta_0} \mathbf{E}_0 \mathbf{s}_0 + \mathbf{E}_0 \sum_{i = 1}^{U} \sqrt{\beta_i} \mathbf{s}_i + \mathbf{F}_0 \mathbf{s} +\mathbf{z}_0.
\end{equation}
{The derivation is provided in} Appendix \ref{Gyreformulated}.
Therefore, the SNR for detecting the HM user's signal at the HM side is expressed as follows,
\begin{equation}\label{snr0}
    \gamma_0 = \frac{\beta_0 \rho_{\mathrm{T}} \omega_{\mathrm{E}}}{(1-\beta_0) \rho_{\mathrm{T}} \omega_{\mathrm{E}}  + \rho_{\mathrm{T}} \omega_{\mathrm{F}} +  \omega_0 },
\end{equation}
where $\omega_{\mathrm{E}}$, $\omega_{\mathrm{F}}$ and $\omega_0$ are given by,
\begin{equation}\label{ww}
    \omega_{\mathrm{E}}  = \sum_{i = 1}^{NM} \frac{\left| \delta_{{\mathrm{E}},i} \right|^2}{NM},    \quad
      \omega_{\mathrm{F}}  = \sum_{i = 1}^{NM} \frac{\left| \delta_{{\mathrm{F}},i} \right|^2}{NM},    \quad
    \omega_0  = \sum_{i = 1}^{NM} \frac{\left| \delta_{0,i} \right|^2}{NM}.
\end{equation}
In \eqref{ww}, the elements $\delta_{0,i}$ are found as shown in \eqref{delta0} and $\delta_{{\mathrm{E},i}}$ and $\delta_{{\mathrm{F},i}}$ are respectively, the elements of $\mathbf{\Delta_{E}}$ and $\mathbf{\Delta_{F}}$ expressed as,
\begin{align}
     \delta_{{\mathrm{E},i}} &= \delta_{0,i} \sum_{a=1}^A v_a \lambda_{a,i} , \nonumber \\
    \delta_{{\mathrm{F},i}} &= \delta_{0,i} \sum_{a=1}^A v_a \mu_{a,i},
\end{align}
where $\mu_{a,i}$ are the elements of $\mathbf{D}_{{\mathrm{I}},a}$. {The derivation is provided in} Appendix \ref{snr00}.

The expression in \eqref{snr0} evaluates the detection SNR of the HM user's signal when the system is unaware of the presence of fractional Doppler. In this case, the parameters characterizing the fractional Doppler $\kappa_{j,0,q}$ and $N_{j,q}$ are unknown to the HM receiver, which can therefore only decode the expected part of the signal as shown in equation \eqref{y0q}.

\subsection{Detection with Full Knowledge of Fractional Doppler}
In this section, we assume the receiver is aware of the existence of fractional Doppler and has full knowledge of the HM channel matrix and fractional Doppler parameters, i.e., $\kappa_p$ and $N_p$, for all paths. In this case, the overall channel matrix $\mathbf{H}_0$ is applied at the detection and the received signal can be rewritten as follows,
\begin{align}\label{y00q}
    \mathbf{y}_{00} &= \sqrt{\beta_0} \mathbf{H}_0 \mathbf{s}_0 + \mathbf{H}_0 
 \sum_{i = 1}^{U} \sqrt{\beta_i} \mathbf{s}_{i}  + \mathbf{w}_0.
\end{align}
At the HM side, the receiver utilizes the following detection matrix,
\begin{equation}
    \mathbf{G}_{00} = \left( {\mathbf{H}}_0^\mathrm{H} \mathbf{H}_0 + \rho \mathbf{I}_{NM}\right)^{-1} {\mathbf{H}}_0^\mathrm{H}.
\end{equation}
The equalized signal is expressed as follows,
\begin{equation}
    \mathbf{G}_{00} y_{00} = \sqrt{\beta_0} \mathbf{M}_0 \mathbf{s}_0 + \mathbf{M}_0 \sum_{i = 1}^{U} \mathbf{s}_i + \mathbf{G}_{00} \mathbf{w}_0,
\end{equation}
where the matrix $\mathbf{M}_0$ is expressed as,
\begin{equation}\label{m0}
    \mathbf{M}_0  = \mathbf{\Psi}^{\mathrm{H}} \mathbf{D}_{00} \mathbf{\Psi}.
\end{equation}
and the diagonal matrix $\mathbf{D}_{00}$ is determined by the decomposition of $\mathbf{G}_{00} \mathbf{H}_0$. Therefore, the SNR at the HM user is expressed as,
\begin{equation}
    \gamma_{00} = \frac{\beta_0 \rho_{\mathrm{T}} \omega_{\mathrm{M}}}{ (1 - \beta_0) \rho_{\mathrm{T}} \omega_{\mathrm{M}} + \omega_{00}},
\end{equation}
where $\omega_{\mathrm{M}}$ and $\omega_{00}$ are given by,
\begin{align}
    \omega_{\mathrm{M}} &= \frac{1}{NM} \sum_{ i = 1}^{NM} |\lambda_{M,i}|^2 \nonumber \\
     \omega_{00} &= \frac{1}{NM} \sum_{ i = 1}^{NM} |\lambda_{00,i}|^2,
\end{align}
where $\lambda_{M,i}$ and $\lambda_{00,i}$ are expressed as,
\begin{align}
    \lambda_{M,i} &= \frac{|\sum_{a = 1}^A v_a \zeta_{a,i}|^2}{|\sum_{a = 1}^A v_a \zeta_{a,i}|^2 + \rho}, \nonumber \\
     \lambda_{00,i} &= \frac{\sum_{a = 1}^A v_a^{*} \zeta_{a,i}^*}{|\sum_{a = 1}^A v_a \zeta_{a,i}|^2 + \rho},
\end{align}
$\zeta_{a,i}$ being the elements of $\mathbf{D}_{a,0}$.\\
{The purpose of the full-detection scheme is to provide a deeper understanding of the impact of fractional Doppler on OTFS-NOMA performance. By assuming that the fractional-Doppler parameters are known at the receiver, the resulting performance reflects the achievable gains when the corresponding channel information is available. This allows us to quantify the performance loss associated with fractional Doppler. Moreover, we can have insights on how much loss can be mitigated when the fractional Doppler parameters are accurately estimated.}

\subsection{Signals Detection at LM Users}
{As discussed earlier, each LM user first detects and removes the HM user's OTFS signal before decoding its own information signal. LM users do not require multiple stages of SIC since they occupy orthogonal subcarriers.}
First, each LM user detects the HM user's signal using the following matrix,
\begin{equation}\label{gu}
    \mathbf{G}_u = \left( {\mathbf{H}}_u^\mathrm{H} \mathbf{H}_u + \rho \mathbf{I}_{NM}  \right)^{-1} {\mathbf{H}}_u^\mathrm{H}.
\end{equation}
The channel matrix $\mathbf{H}_u$ can be diagonalized as follows,
\begin{equation}
    \mathbf{H}_u = \mathbf{\Psi}^\mathrm{H} \mathbf{D}_u \mathbf{\Psi},
\end{equation}
where $\mathbf{D}_u$ is a diagonal matrix comprising the elements $\sum_{a = 1}^A \lambda_{u,a,i}, 1 \le i \le NM$. Thus, the matrix in \eqref{gu} can be equivalently expressed as follows,
\begin{equation}
    \mathbf{G}_u = \mathbf{\Psi}^{\mathrm{H}} \mathbf{\Delta}_u \mathbf{\Psi},
\end{equation}
where $\mathbf{\Delta}_u$ is a diagonal matrix consisting of elements,
\begin{equation}
    \delta_{u,i} = \frac{\sum_{a = 1}^A   \left({v_a} \lambda_{u,a,i} \right)^*}{ \left |\sum_{a = 1}^A  v_{a} \lambda_{u,a,i} \right|^2 + \rho }, 1 \le i \le NM.
\end{equation}
Therefore, after applying the equalizer to the received signal in \eqref{yu}, we get the following signal,
\begin{equation}
    \mathbf{G}_u y_u = \sqrt{\beta_0} \mathbf{G}_u \mathbf{H}_u \mathbf{s}_0 + \mathbf{G}_u \mathbf{H}_u \sum_{i = 1}^{U}\sqrt{\beta_i} \mathbf{s}_i + \mathbf{G}_u \mathbf{w}_u.
\end{equation}
Following the same steps to derive \eqref{snr0} and assuming perfect detection, the detection SNR of the HM user at LM user $u$ is expressed as follows,
\begin{equation}\label{snr0u}
    \gamma_{0,u} = \frac{\beta_0 \rho_{\mathrm{T}} \omega_{T}}{ \sum_{i = 1}^{U} \beta_i \rho_{\mathrm{T}}  \omega_{T} + \omega_\mathrm{U}}, 
\end{equation}
where $\omega_T$ and $\omega_U$ are given by,
\begin{align}
    \omega_\mathrm{T} &= \frac{1}{NM} \sum_{i = 1}^{NM} \left | \delta_{u,i} \right|^2 \left|\sum_{a = 1}^A   v_a\lambda_{u,a,i} \right|^2, \\
    \omega_\mathrm{U} &= \frac{1}{NM} \sum_{i = 1}^{NM} \left | \delta_{u,i} \right|^2.
\end{align}

We denote by $H_{u,a}[n,m]$ the channel gain between user $u$ and antenna $a$ expressed in the TF domain. After decoding and subtracting the HM user's signal from its received signal, each LM user $u, 1 \le u \le U$ observes the following received signal in the TF domain at instant $nT$ and subcarrier $m\Delta f$,
\begin{align}\label{yu1}
    y_u[n,m] &= \sum_{a = 1}^A v_{a} {H}_{u,a}[n,m] \sum_{i = 1}^{U}  \beta_i s_{i}[n,m] + w_u[n,m],
\end{align}
where $w_u[n,m]$ is the noise term.
{Herein, perfect SIC is assumed in order to isolate the impact of fractional Doppler and IDI on the performance of OTFS-NOMA HetNets. In practical implementations, imperfect SIC results in residual interference that further degrades the detection performance of LM users} \cite{sabuj2020cognitive}.
Considering the LM users are not subject to Doppler shifts, the channel gain is time-independent and is given by,
\begin{equation}
    H_{a,u}[m] = \sum_{j=1}^{P_u} \alpha_{j,u} e^{j2\pi \frac{l_{{j},u} m}{M}},
\end{equation}
where $l_{j,u}$ is the delay tap of path $j$ expressed as in \eqref{taps}. 

Since LM users signals in the same picocell are orthogonal in the frequency domain as in \eqref{suq}, user $u$ considers only the received signal on its dedicated subchannel $m$ and disregards the other subchannels. {Since LM users experience flat-fading channels, we assume that the coherence bandwidth of the LM user's channel is larger than the bandwidth of each subchannel.}
As a result, we can use a single-tap equalizer to detect the LM user's signal as follows,
\begin{align}
    \frac{y_u[n,u-1]}{H_u[u-1]} = \sqrt{\beta_u} s_u[n,u-1] 
   + \frac{w_u[n,u-1]}{H_u[u-1]},
\end{align}
where $H_u[u-1] = \sum_{a = 1}^A v_a H_{u,a}[u-1]$.
{The detection procedure at the LM users follows the conventional SIC principle adopted in power-domain NOMA systems.}
 The SINR of LM user's signals detection is time-invariant and is given by,
\begin{equation}\label{gammau}
    \gamma_u = \beta_u \rho_{\mathrm{T}} \left| H_u[u-1] \right|^2.
\end{equation}
\textbf{Remark:}
We can adapt the NOMA allocation factors in order to guarantee an acceptable performance for all users, including the HM user that suffers from IDI, which cannot be mitigated by increasing the OTFS resolution to avoid communication latency.

\section{Optimization of NOMA Power Factors}\label{optS}

When adopting NOMA scheme, the quality of detection of the HM user's signal is crucial at both HM and LM sides with immediate effect on their performance. Therefore, we aim to increase the HM signal's detection SNRs at both the HM and LM side. 
These SNRs are increasing with the NOMA factor $\beta_0$ as can be seen in \eqref{snr0} and \eqref{snr0u}. 
Thus, increasing this parameter improves the detection quality. However, this results in a decreased allocated power to LM users. As a result, our objective in this section is to optimize the performance of the OTFS-NOMA system by optimizing the HM user's transmit power at the PBS while maintaining a minimum QoS requirement at LM users side. The performance of LM users in each PBS is evaluated through their instantaneous DL sum-rate expressed as follows,
\begin{align}\label{rs}
    R_{\mathrm{S}} &= \sum_{u = 1}^{U} \log_2 \left(1 + \gamma_u  \right) \nonumber \\
    &= \sum_{u = 1}^{U} \log_2 \left(1 +  \beta_u a_u \right),
\end{align}
where the constant $a_u$ is obtained as,
\begin{equation}
    a_u = \rho_{\mathrm{T}} \left| \sum_{a = 1}^A v_a H_{u,a}[u-1] \right|^2.
\end{equation} 
We express the considered optimization problem as follows,
\begin{equation}\label{pb}
\begin{aligned}
    \max_{\beta_{u},u = 0, \ldots, U} \quad & \beta_{0}  + \lambda_{\mathrm{L}} \sum_{u = 1}^{U} \log_2 \left(1 + \beta_u a_u \right), \\
    \textrm{s.t.} \quad & {\rm{C1:}} \; \sum_{u = 1}^{U} \log_2 \left(1 + \beta_u a_u  \right) \ge R_{\min} \\
     & {\rm{C2:}} \;\sum_{u = 0}^{U} \beta_{u} \le 1 , \\
      & {\rm{C3:}} \; \beta_{u} \ge 0 , \quad \forall u = 0, \ldots, U,
\end{aligned}
\end{equation}
where $R_{\min}$ is the minimum required LM sum-rate and $\lambda_{\mathrm{L}} > 0$ is a Lasso regularization parameter used to determine the level of fairness between different users. 
In problem \eqref{pb}, the constraint (C1) is used to guarantee a minimum QoS of LM users within the picocell expressed as the LM users sum-rate. The constraint (C2) is set to ensure the allocated power to the MUs does not exceed the maximum budget at the BS. \\
{The optimization variables are coupled through both the objective function and the problem constraints. In particular, increasing the HM user power allocation factor $\beta_0$ reduces the power available to LM users through the total power constraint. Moreover, the power allocation factors of LM users jointly determine the achievable sum-rate and therefore affect the feasibility of the minimum sum-rate requirement.}\\
The value of $\lambda_{\mathrm{L}}$ is a tuning parameter to determine the level of fairness between different users. If $\lambda_{\mathrm{L}} = 0$, maximizing the HM user's detection performance is prioritized, while LM users may not receive data messages from the PBS. Increasing $\lambda_{\mathrm{L}}$ to a strictly positive value maximizes the sum-rate of LM users, and thus, more LM users get service from the PBS.\\
The optimization problem in \eqref{pb} is convex (the objective function and the constraints are sums of concave functions) and can be solved using Lagrange multipliers \cite{boyd2004convex}. 
We relax the problem by removing the positive sign constraint (C3).
By relaxing (C1) and (C2), the corresponding Lagrangian can be obtained as follows,
\begin{align}
    L \left(\beta_u,\theta_1,\theta_2 \right) &=  -\beta_0 - \lambda_{\mathrm{L}} \sum_{u = 1}^{U} \log_2 \left(1 + a_u \beta_u  \right)  \nonumber \\
    &+ \theta_1 \left( R_{\min} - \sum_{u = 1}^{U} \log_2 (1 + a_u \beta_u) \right) \nonumber \\
    &+ \theta_2 \left( \sum_{u = 0}^{U} \beta_u - 1 \right),
\end{align}
where $\theta_1, \theta_2 \ge 0$ are Lagrange multipliers.
The optimal solution for LM users, i.e., $u = 1, \ldots, U$ is,
\begin{equation}\label{pu}
    \beta_u = \max \left( X_{U} , \frac{\lambda_{\mathrm{L}}}{\log(2)} \right) - \frac{1}{a_u},
\end{equation}
where $X_{U}$ can be computed as follows,
\begin{equation}\label{XU}
    \log_2(X_U) = \frac{R_{\min} - \sum_{u = 1}^U \log_2(a_u)}{U}, \quad U \ge 1.
\end{equation}
Finally, the optimal power allocation for the HM user is as calculated as,
\begin{equation}
    \beta_0 = 1 - \sum_{u = 1}^{U} \beta_u.
\end{equation}
{The derivation is provided in }Appendix \ref{optimization}.

The value of $\beta_u$ in \eqref{pu}, which is the solution of the relaxed problem, can be negative depending on the system parameters, mainly the number of LM users in the cell $U$ and their channel conditions. Therefore, the optimal solution of the original problem in \eqref{pb} (with active condition (C3)) has zero transmission power for some LM users, i.e., there exists {at least} a user $u$ {such that} $\beta_u = 0$. This implies that these LM users do not receive data signals from the PBS. 
The LM users with the weakest channel gains have less contribution to the sum-rate in \eqref{rs} than the LM users with better channel conditions which have better chances to increase the sum-rate to reach $R_{\min}$ with less power consumption. Thus, these weakest users will be assigned zero transmission powers. 
Therefore, if we consider the factors $a_u$ to be sorted such that $a_1 > \ldots > a_U$, the optimal solution for the LM users is expressed as,
\begin{align}\label{puopt}
    \beta_u^* = 
    \begin{cases}
        & \max \left(X_{U^*},\frac{\lambda_{\mathrm{L}}}{\log(2)} \right) - {1}/{a_u}, \quad u = 1, \ldots, U^*, \\
        & 0, \qquad \qquad \qquad \qquad \qquad u = U^*+1, \ldots, U, \\
        & 1 - \sum_{u = 1}^{U^*} \beta_u^*, \qquad \qquad \quad u = 0,
    \end{cases}
\end{align}
which is achieved with the highest number of users $U^* \le U$ that satisfies, $\beta_u^* \ge 0, u = 1, \ldots, U^*$ and $\sum_{u = 1}^{U^*} \beta_u^* = 1$. The number $U^*$ is determined by iteratively removing the users with negative $\beta_u$ expressed in \eqref{pu} and redistributing the power allocation between the remaining users until all $\beta_u \ge 0$ for all LM users. The allocation process is further explained in Algorithm \ref{alg}.

The solution of this problem results in a zero transmit power to multiple LM users depending on the system parameters and on the choice of $\lambda_{\mathrm{L}}$. Decreasing $\lambda_{\mathrm{L}}$ gives priority to the HM user power factor.
This might lead to a high number of LM users not being served as long as the minimum sum-rate is achieved.
In order to partially address this limitation, we can increase $\lambda_{\mathrm{L}}$ and penalize the objective function in case of a decrease in the sum-rate. 

{The proposed optimization setup aims at maximizing the HM user's SNR without depending on the OTFS parameters. Thus, the resource allocation framework can ensure fairness and efficiency and does not depend on the system's knowledge of fractional Doppler and its exact value.} 
{Algorithm 1 is initialized with the complete set of LM users and the corresponding channel coefficients. The algorithm computes the initial power allocation factors and identifies users assigned negative power values. These users are removed from the active LM user set. Then, the allocation is recomputed iteratively until all power allocation factors are greater than zero. Finally, the power-allocation factor of the HM user is obtained from the remaining available power budget.}
\begin{algorithm}
\caption{Proposed Power Allocation Scheme}\label{alg}
\begin{algorithmic}
\State \textbf{Input:} $U, a_u, R_{\min}, \lambda_{\mathrm{L}}$
\vspace{3pt}
\State Sort LM users such that $a_1 > a_2 > \cdots > a_U$
\vspace{3pt}
\State Initialize the active LM-user set: $\mathcal{U}_{\mathrm{a}}={1,\ldots,U}$
\vspace{3pt}
\State Set $U_1=U$
\vspace{3pt}

\While{$\exists \; u \in \mathcal{U}_{\mathrm{a}}$ such that $\beta_u<0$}
\vspace{5pt}

\State Compute
$
X_{U_1} =
2^{\frac{R_{\min}-\sum_{u \in \mathcal{U}_{\mathrm{a}}}\log_2(a_u)}
{U_1}}
$
\vspace{5pt}

\State Compute
$
\beta_u =
\max(X_{U_1},\lambda_{\mathrm{L}}/\log(2))
-\frac{1}{a_u},
\quad \forall u \in \mathcal{U}_{\mathrm{a}}
$
\vspace{5pt}

\State Remove users with negative allocation:
$
\mathcal{U}_{\mathrm{a}}
=
\mathcal{U}_{\mathrm{a}}
\setminus
{u:\beta_u<0}
$
\vspace{5pt}

\State Update $U_1=|\mathcal{U}_{\mathrm{a}}|$
\vspace{5pt}

\EndWhile

\vspace{3pt}
\State Set $U^*=U_1$
\vspace{3pt}

\State Compute the final LM-user power-allocation factors:
$
\beta_u^*=
\max(X_{U^*},\lambda_{\mathrm{L}}/\log(2))
-\frac{1}{a_u},
\quad \forall u\in\mathcal{U}_{\mathrm{a}}
$
\vspace{3pt}

\State Set $\beta_u^*=0,\quad \forall u\notin\mathcal{U}_{\mathrm{a}}$
\vspace{3pt}

\State Compute the HM-user power-allocation factor:
$
\beta_0^*
=
1-\sum_{u\in\mathcal{U}_{\mathrm{a}}}\beta_u^*
$
\vspace{3pt}

\State \textbf{Output:} $\beta_0^*,\beta_1^*,\ldots,\beta_U^*$

\end{algorithmic}
\end{algorithm}

{The algorithm guarantees that all active users satisfy the constraints of problem}\eqref{pb}. {Users assigned zero power are excluded from transmission, while the remaining users receive power allocations that satisfy both the minimum sum-rate requirement and the total power budget.}\\
{The computational complexity of the proposed framework is dominated by the MMSE detector, which requires the inversion of the channel matrix. In contrast, the proposed power allocation algorithm involves only iterative updates of the active set of LM users and therefore has a significantly lower computational cost.}
\section{Simulation Results}
In this section, we conduct simulations in order to explore the performance of the proposed system. We follow the law of total expectations when computing the rates and the law of total probabilities when computing the probabilities. {Analyzing system performance in terms of spectral efficiency is essential since it helps show the ability and performance of the system in maximizing data transmission over limited bandwidth. As a result, we can optimize resource utilization and improve the overall system throughput, which are critical aspects in current and future wireless communication networks.}
We consider the following simulation parameters unless otherwise specified: $N_0 = 5$, $\rho_{\mathrm{T}} = 25 \; \mathrm{dB}$, $U_{\max} = 8$, $P_0^q=4$. 
The value $P_0^q=4$ {is consistent with sparse delay-Doppler channel models commonly adopted in OTFS studies.}
The number $U_{\max}$ {reflects the maximum number of LM users that can be simultaneously served by the considered picocell.}
{The delay taps are also commonly adopted in OTFS studies and are the same as in} \cite{ding2019otfs}. The maximum Doppler shift is for a vehicle speed of 400 km/h. The Doppler shifts are randomly selected according to Jake's model. The number of LM users is uniformly distributed between 1 and $U_{\max}$. If not specified, we assume $\beta_0 = 0.75$. The antenna factors are set to $v_a = 1/\sqrt{A}, \forall 1 \le a \le A$. {In our study, equal weights are assigned to all PBS antenna elements in order to focus on the impact of fractional Doppler and NOMA power allocation rather than antenna-weight optimization. Preliminary investigations showed that varying the antenna weights has a negligible effect on the average spectral efficiency considered in this work, and therefore uniform weighting is adopted for simplicity.}
  \begin{figure}[!t]
	\centering 
	\captionsetup{justification=centering,margin=1cm}
	\includegraphics[width=3.5in]{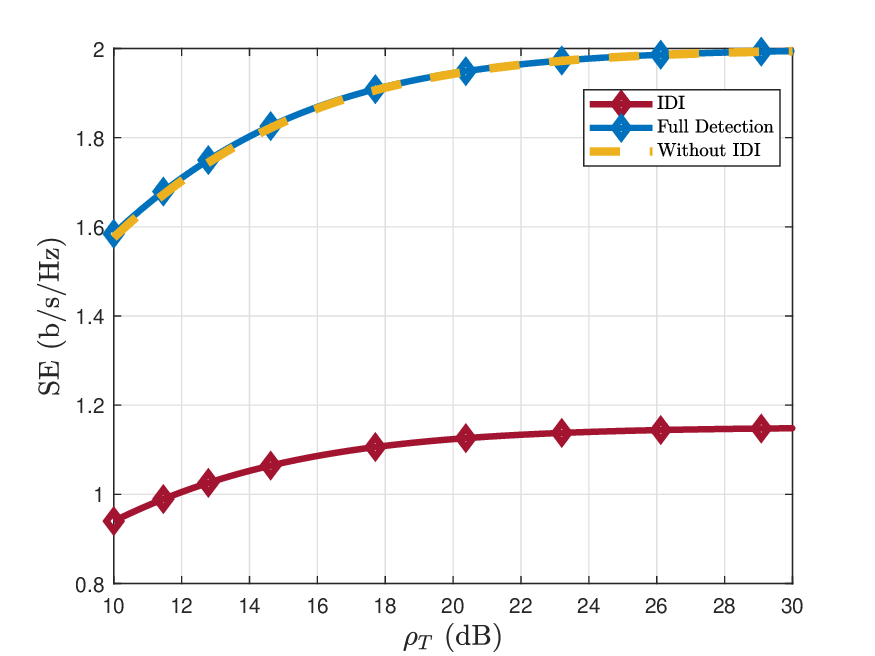}
	\caption{Spectral efficiency comparison with respect to the transmit SNR of the ideal system with no IDI, the real system with IDI and a scheme where full detection is performed.}
	\label{kappa}
\end{figure}
  \begin{figure}[!t]
	\centering 
	\captionsetup{justification=centering,margin=1cm}
	\includegraphics[width=3.5in]{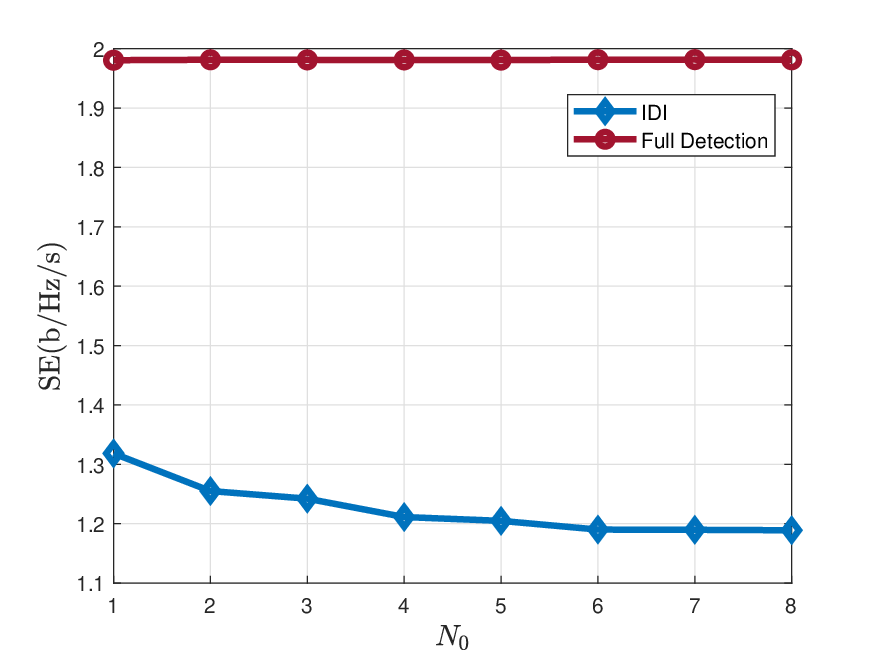}
	\caption{Spectral efficiency of HM user when full detection and IDI schemes are applied with respect to the number of subpaths.}
	\label{np}
\end{figure}

First, we investigate the effect of the fractional Doppler on the actual (practical) system performance. We explore the performance of both the ideal (without IDI) and real (with IDI) performance corresponding to the assumption of absence and presence of fractional Doppler, respectively. Taking the fractional Doppler into consideration implies that the IDI interference is taken into account during system evaluation. {The "Without IDI" curve serves as a benchmark corresponding to the conventional OTFS-NOMA model where fractional Doppler is neglected.}
To this end, we plot the HM rate versus the transmit SNR as shown in Fig. \ref{kappa}. The performance of the ideal scheme serves, in reality, as an upper limit to the real performance of the OTFS-NOMA system in high mobility scenarios since the IDI is an additional interference to the signal. 
The observed performance gap is significant and increases with the transmit SNR. At $\rho_{\mathrm{T}} = 25 \; \mathrm{dB}$, the spectral efficiency (SE) gap grows to 0.8 b/s/Hz, which leads to a significant throughput degradation. As a result, we can conclude that not taking into account the IDI interference leads to incorrect assessment of the OTFS-NOMA system and, thus, an unsuitable system setting. 
Moreover, we notice that the full detection scheme achieves the same performance as the case of absence of IDI. This means that, if the channel parameters are known at the receiver, full detection can be applied and the performance of the system can be significantly improved.

Then, we assess the impact of channel information in terms of the number of subpaths $N_0$ on the system performance. In Fig. \ref{np}, we plot the spectral efficiency at the HM side versus the number of subpaths $N_0$ for both cases of IDI presence and full detection. First, we notice that when full detection is applied at the HM receiver, the system performance remains constant as the number of subpaths $N_0$ increases. This confirms the previous result about the similar performance between full detection and absence of IDI. Second, in the presence of IDI, the spectral efficiency at the HM side slightly decreases with the increase of $N_0$ and becomes constant at high values of $N_0$. $N_0 = 1$ corresponds to the best-case scenario and an upper limit to the performance assuming the existence of fractional Doppler (In general, $N_0 = 0$ is best-case scenario indicating the absence of fractional Doppler). From Fig. \ref{np}, we observe that starting from $N_0 = 5$, the performance remains constant, and therefore, $N_0 = 5$ is the smaller number of subpaths corresponding to the worst-case scenario. 

  \begin{figure}[!h]
	\centering 
	\captionsetup{justification = centering,margin = 1cm}
	\includegraphics[width = 3.5in]{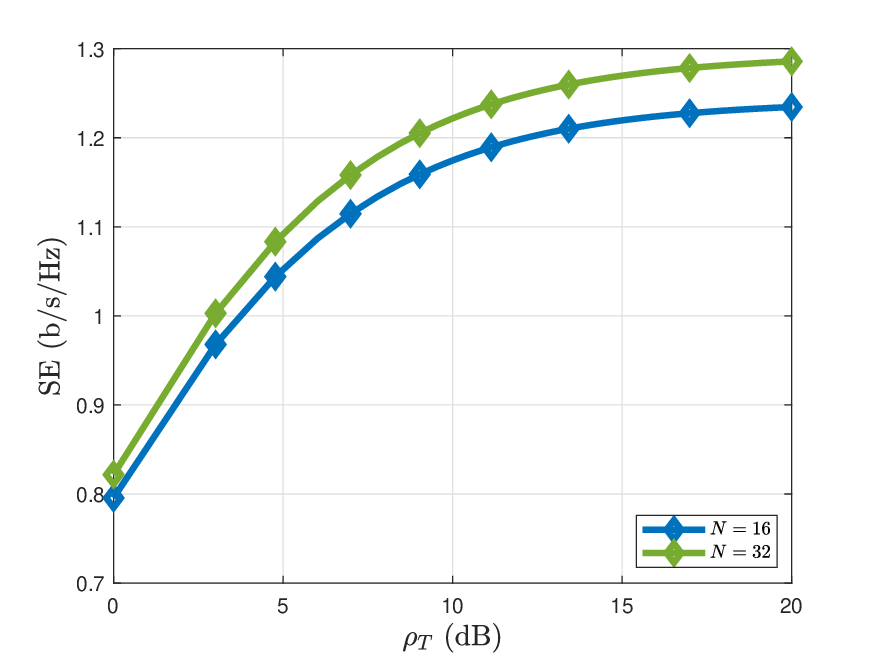}
	\caption{{SE of HM user for different values of $N$.}}
	\label{difN}
\end{figure}
\begin{figure}
     \centering
     \captionsetup{justification=centering,margin=0.5cm}
     \begin{subfigure}[b]{0.5\textwidth}
         \centering
         \captionsetup{justification=centering,margin=1cm}
         \includegraphics[width=3.1in]{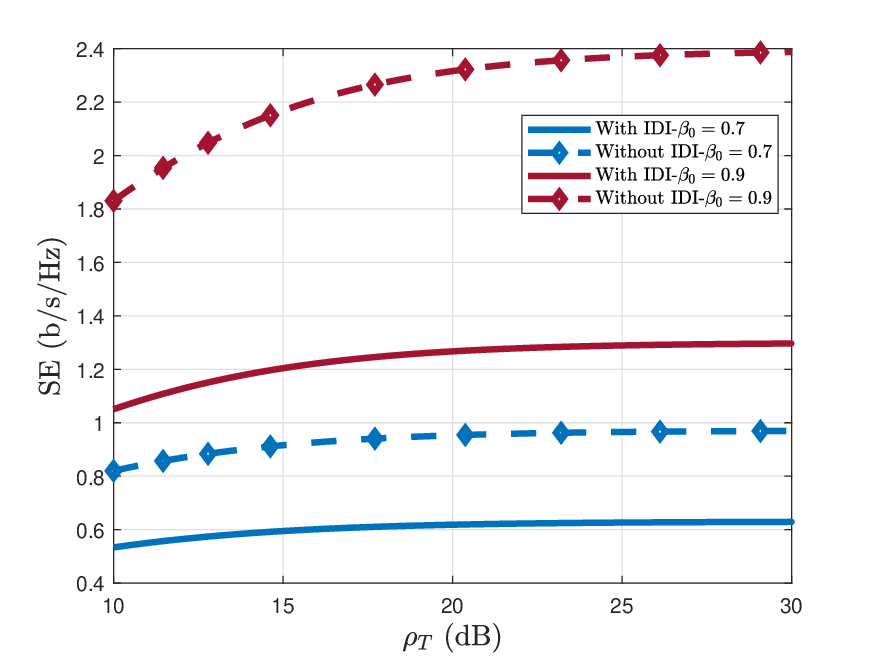}
         \caption{Spectral efficiency at HM side.}
         \label{hmse}
     \end{subfigure}
     \hfill
     \begin{subfigure}[b]{0.5\textwidth}
         \centering
         \captionsetup{justification=centering,margin=1cm}
         \includegraphics[width=3.1in]{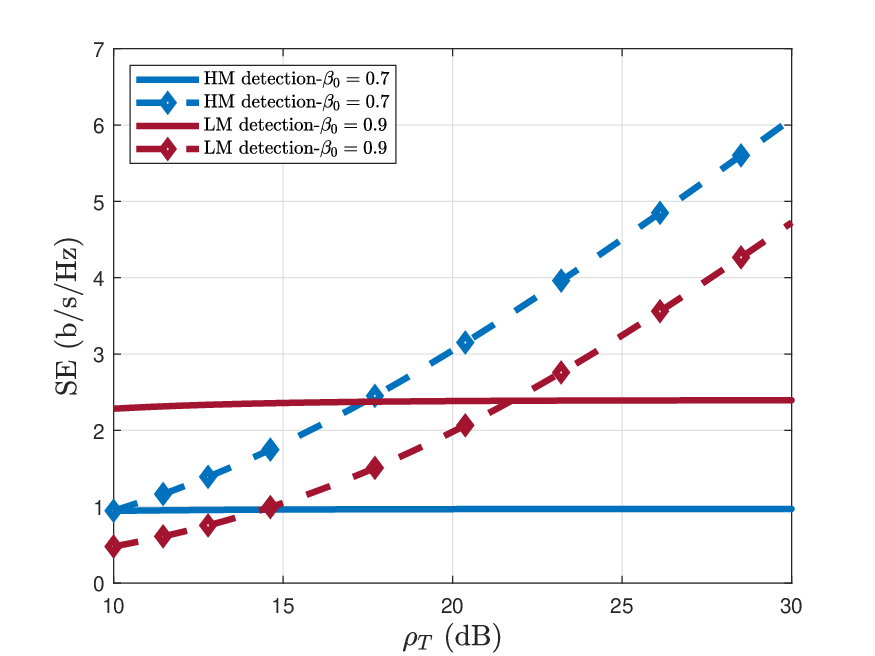}
         \caption{Spectral efficiency at LM side.}
         \label{lmse}
     \end{subfigure}
        \caption{Spectral efficiency for different NOMA allocation factors with respect to various transmit SNR and $\beta_u$ values.}
        \label{betas}
\end{figure}

  {In Fig. 4, we plot the spectral efficiency of HM user versus the SNR for different values of $N$ for the same parameters of fractional Doppler, $N_0$. We notice that a higher number of Doppler bins results in improved performance on the HM side. This is justified by the fact that a greater value of $N$ results in a higher Doppler resolution. Thus, the effect of fractional Doppler decreases. Nonetheless, increasing $N$ results in longer frame durations and transmission latency.}

Next, in Fig. \ref{betas}, we investigate the effect of the NOMA power distribution between users on the HM and LM users' performance. We plot the spectral efficiency versus the transmit SNR for two different values of $\beta_0$ in the considered picocell and for the two cases of IDI presence and absence. According to Fig. \ref{hmse}, the system achieves remarkably better performance for the HM users' DL signal when higher power allocation is considered. Although this difference is more significant when IDI is not considered and can reach 1.4 b/s/Hz, it is also impactful when IDI is taken into account and is around 0.8 b/s/Hz when $\rho_{\mathrm{T}} = 25 \; \mathrm{dB}$. On the other hand, in Fig. \ref{lmse}, we plot the HM detection SNR and LM detection SNR at the LM side. We note that allocating more power to the HM user improves the HM detection SNR and therefore, reduces detection errors at the LM side, but at the same time, reduces the LM detection SNR since less power is allocated to LM users. However, it is important to mention that the performance gap between the two values of the allocation factor is smaller than the gap at the HM side, as seen in Fig. \ref{hmse}.

  \begin{figure}[!t]
	\centering 
	\captionsetup{justification = centering,margin = 1cm}
	\includegraphics[width=3.5in]{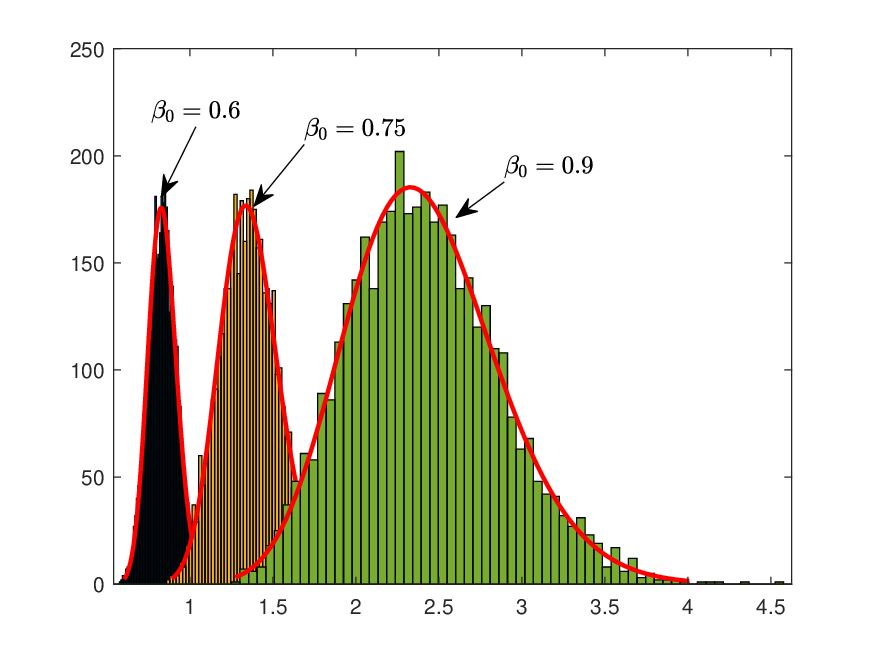}
	\caption{Probability distribution of the SNR at the HM (U is a random variable).}
	\label{pdf}
\end{figure}

We also study the statistical distribution of the SNR at the HM user. In Fig. \ref{pdf}, we plot the probability distribution of the SNR at the HM side in the presence of IDI for three different cases of power allocation. As expected, because of the effect of multipath propagation (including subpaths generating from the fractional Doppler) the SNR values can be fitted to a Gamma distribution with the following parameters: when $\beta_0 = 0.6$, $a = 97.2928$ and $b = 0.0086$. When $\beta_0 = 0.75$, $a = 60.0372$ and $b = 0.0226$. When, $\beta_0 = 0.9$, $a = 28.2183$ and $b = 0.0856$. These values are determined using Matlab distribution fitting functions. We note that for higher power allocation factors, the SNR values are spread across a wider range. Although with higher values of HM power allocation factor $\beta_0$ the SNR can take higher values and improve the average detection SNR of the OTFS signal, the variance of the SNR is higher than the case of lower values of $\beta_0$. This leads to higher uncertainties and less consistent quality of service.

\begin{figure}
     \centering
     \captionsetup{justification=centering,margin=0.5cm}
     \begin{subfigure}[b]{0.5\textwidth}
         \centering
         \captionsetup{justification=centering,margin=1cm}
         \includegraphics[width=3.1in]{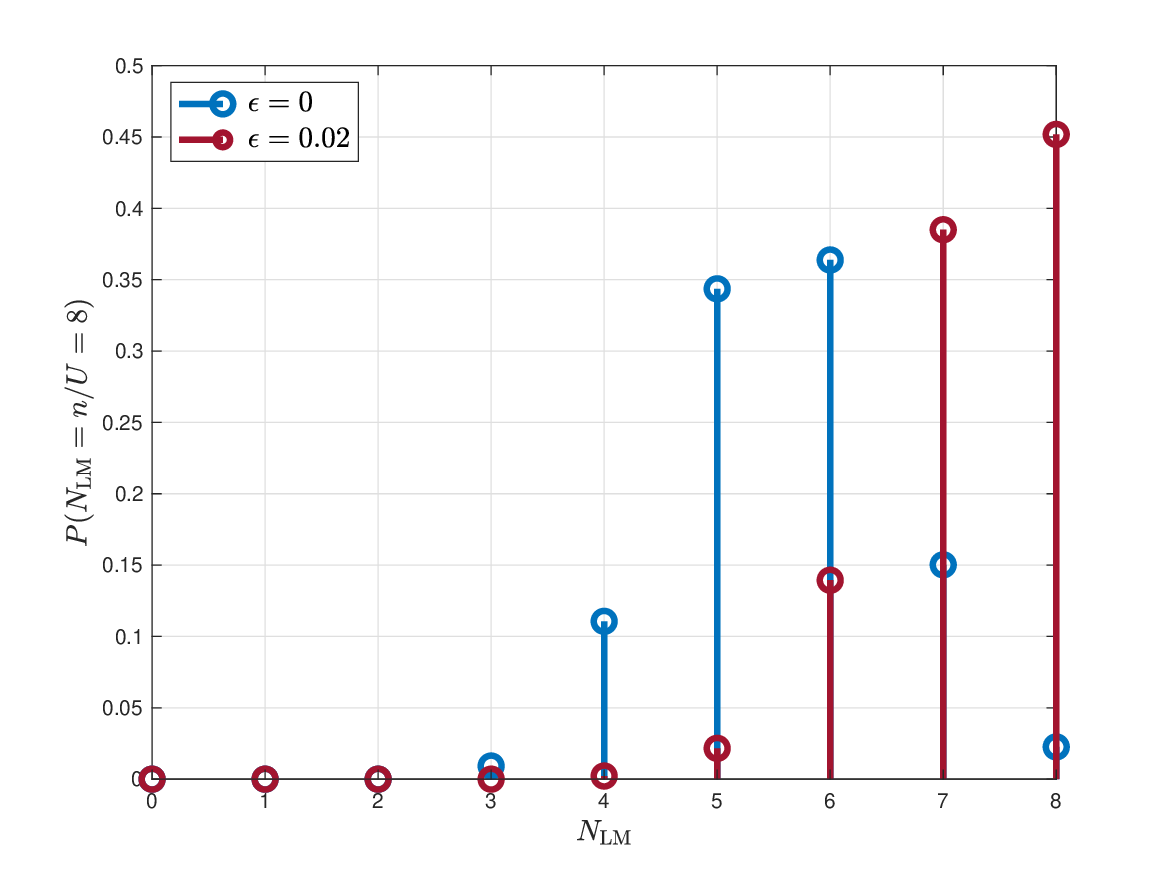}
         \caption{Conditional PDF of the number of LM users served where $U = 8$.}
         \label{cond}
     \end{subfigure}
     \hfill
     \begin{subfigure}[b]{0.5\textwidth}
         \centering
         \captionsetup{justification=centering,margin=1cm}
         \includegraphics[width=3.1in]{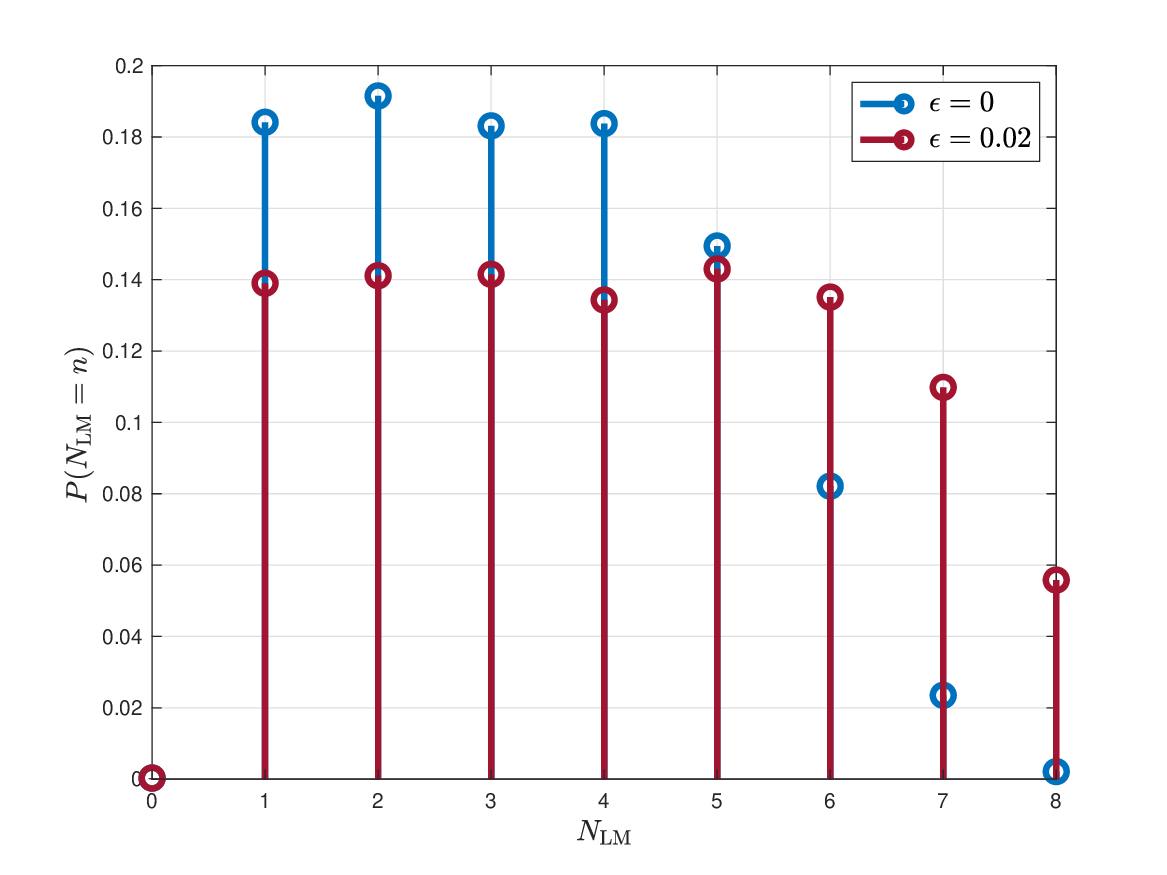}
         \caption{PDF of the number of LM users served where $U$ is a uniform random variable.}
         \label{gen}
     \end{subfigure}
        \caption{PDF of the number of LM users served when $\rho_{\mathrm{T}} = 25 \; \mathrm{dB}$.}
        \label{pdfNU}
\end{figure}
\begin{figure}
     \centering
     \captionsetup{justification=centering,margin=0.5cm}
     \begin{subfigure}[b]{0.5\textwidth}
         \centering
         \captionsetup{justification=centering,margin=1cm}
         \includegraphics[width=3.1in]{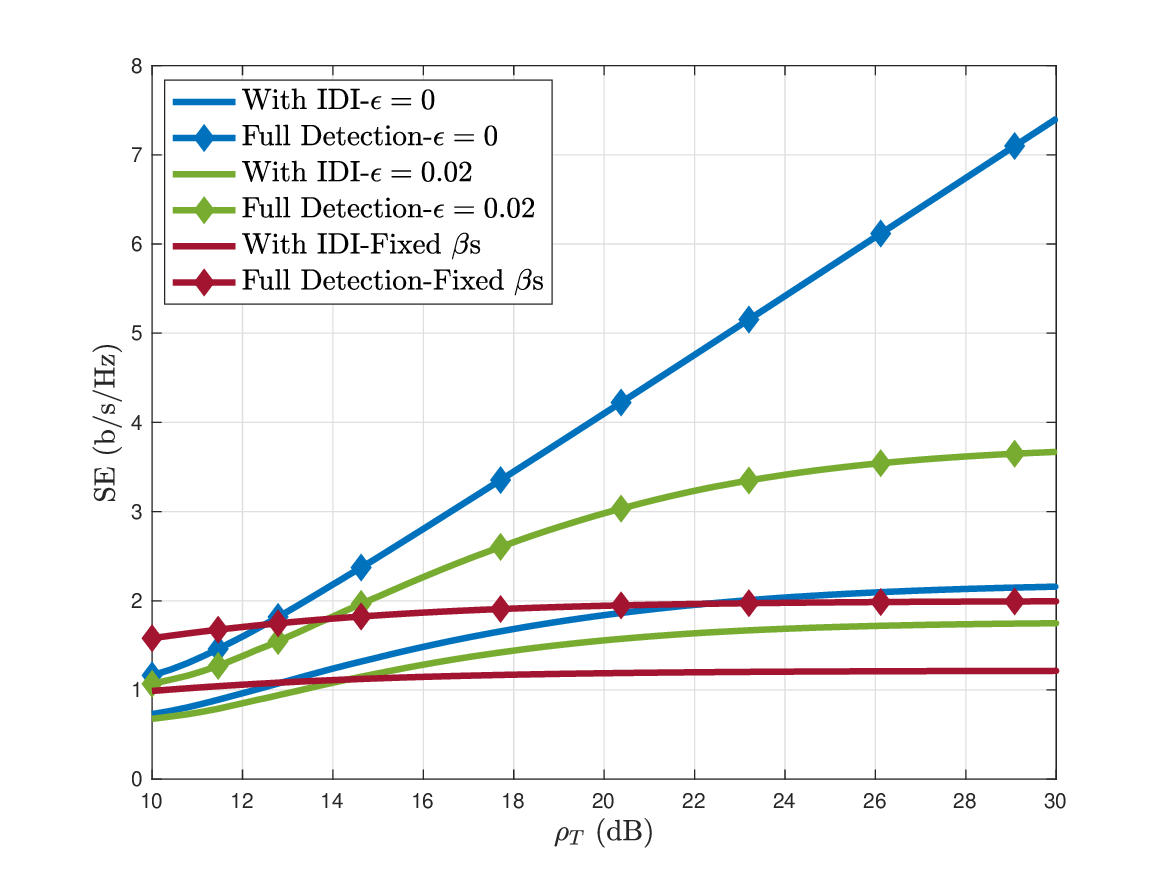}
         \caption{HM rate comparison between different power allocation schemes.}
         \label{hmLasso}
     \end{subfigure}
     \hfill
     \begin{subfigure}[b]{0.5\textwidth}
         \centering
         \captionsetup{justification=centering,margin=1cm}
         \includegraphics[width=3.1in]{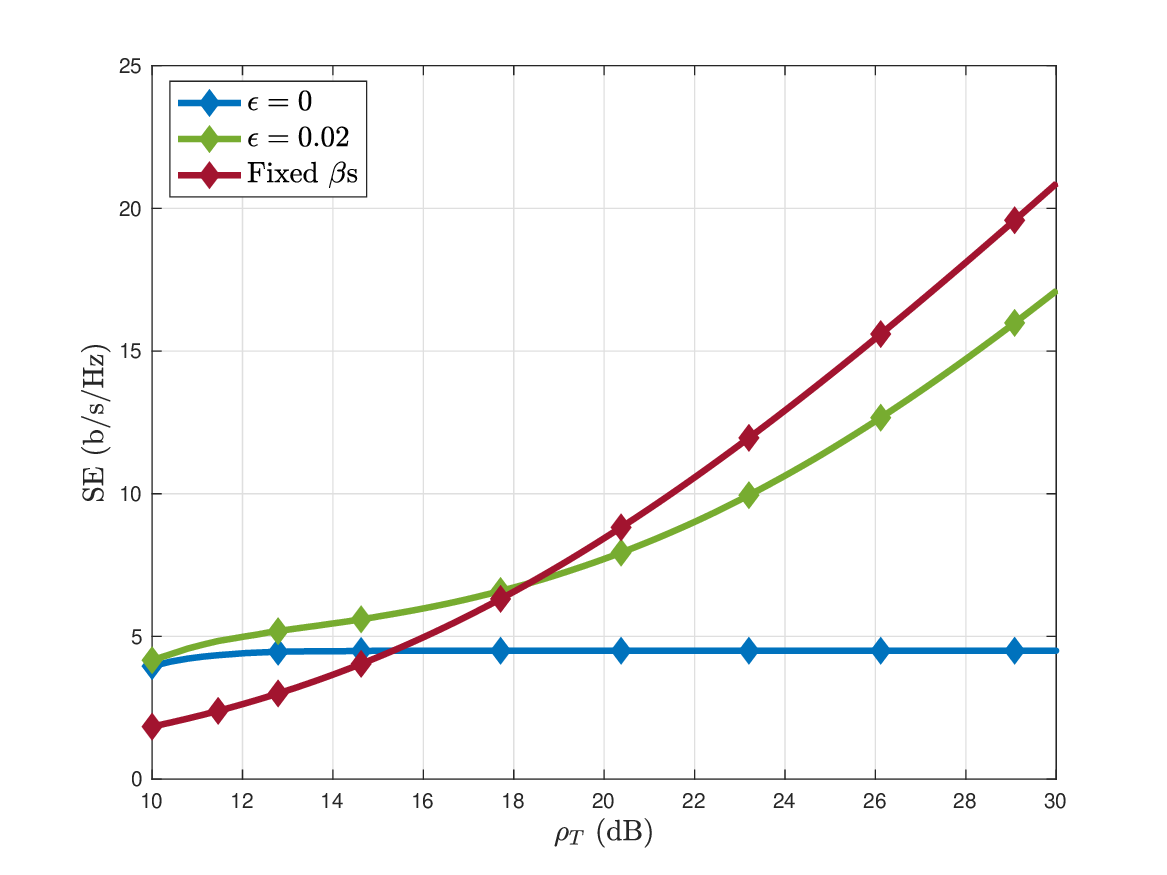}
         \caption{LM sum-rate comparison between different power allocation schemes.}
         \label{lmLasso}
     \end{subfigure}
        \caption{Comparison of users' performance  in terms of spectral efficiency with respect to various transmit SNR values and for different power allocation schemes.}
        \label{lasso}
\end{figure}

We also evaluate the probability distribution of the number of LM users getting service for different values of $\epsilon$ specified in \eqref{epp}, where $\epsilon = 0.02$ corresponds to Lasso optimization scheme. In Fig. \ref{cond}, we plot the conditional probability density function (PDF) of the number of LM users getting service from the PBS. We observe that increasing the value of $\epsilon$ from 0 to 0.02 increases the probability of higher numbers of LM users getting service from the PBS. Fig. \ref{gen} shows the PDF of the number of LM users getting service when $U$ is a random variable. The same behavior can be observed in the case of $\epsilon = 0.02$ where the probability of more LM users getting service increases. Thus, setting $\epsilon = 0$ results in a limited number of LM users receiving data packets from the PBS. Increasing $\epsilon$ improves the probability of all LM users getting service from the PBS. {{The results indicate that increasing the regularization parameter reduces the tendency of the optimization process to allocate resources to only a small subset of LM users. Instead, a larger number of users can be simultaneously served by the PBS, leading to a more balanced distribution of resources among LM users.}}

We further evaluate the performance of the proposed system while considering the optimization scheme presented in Section \ref{optS}. According to the solution in \eqref{puopt}, if $\lambda_{\mathrm{L}} < \log(2) X_{U^*}  $, the Lasso regularization factor $\lambda_{\mathrm{L}}$ does not have any effect on the solution and the solution is the same as in the case of $\lambda_{\mathrm{L}} = 0$. Therefore, in order to reap the benefit of Lasso regularization in improving the sum-rate of LM users, in the simulations, we set $\lambda_{\mathrm{L}}$ as follows,
\begin{equation}\label{epp}
    \lambda_{\mathrm{L}} =\log(2) X_{U^*}  + \epsilon,
\end{equation}
where $\epsilon>0$ is a positive number. This choice forces the condition $\lambda_{\mathrm{L}} > \log(2) X_{U^*}$.\\
We compare the proposed optimization scheme to a fixed power allocation scheme. {The fixed power-allocation schemes are used as benchmark references to evaluate the gains achieved by the proposed optimization framework.}
To this end, in Fig. \ref{lasso}, we plot the spectral efficiency of the HM user and the sum-rate of LM users versus the transmit SNR for different values of $\epsilon$ and $\beta_0$ and considering IDI and full detection. For fixed $\beta$ case, we consider $\beta_0 = 0.75$ and $\beta_u = \left(1-\beta_0 \right) \frac{|H_u|^{-1}}{\sum_{i = 1}^U |H_i|^{-1}}$.
We can observe that for $\epsilon = 0.02$, the LM users' performance improves at the expense of decreased HM user performance for both cases of IDI presence and full detection. The performance gap between the two cases of $\epsilon$ is higher for the full detection case.
Therefore, as confirmed by the previous simulation example, $\epsilon$ can be tuned to improve fairness between users and improve the sum-rate of LM users. Moreover, for high SNR regimes, the optimized scheme provided in this work outperforms the fixed power allocation scheme in terms of HM spectral efficiency. {This optimization framework helps demonstrate the interplay between IDI and NOMA power allocation, without requiring prior information about fractional Doppler parameters.}

  \begin{figure}[!t]
	\centering 
	\captionsetup{justification = centering,margin=1cm}
	\includegraphics[width = 3.5in]{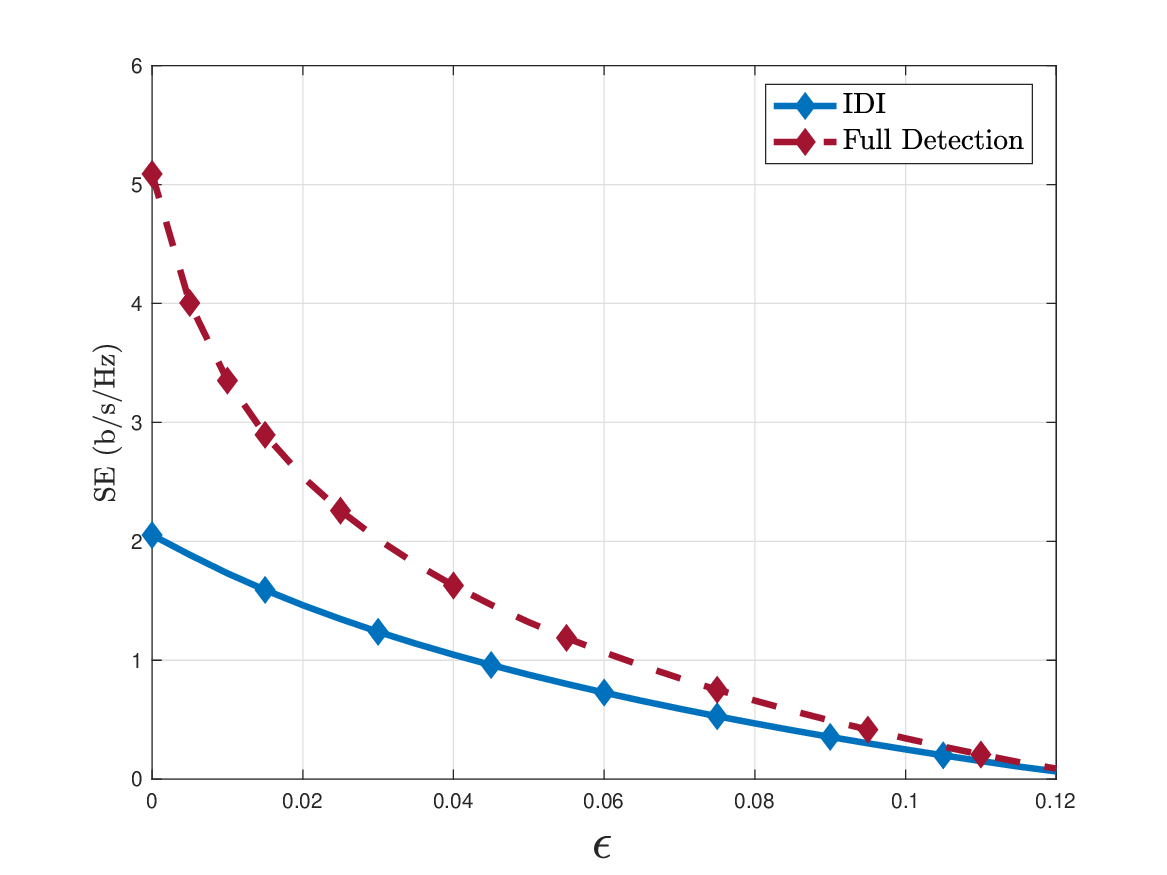}
	\caption{HM spectral efficiency rate versus $\epsilon$ for different HM speeds and $U = 8$.}
	\label{epsilon}
\end{figure}

In our last evaluation, we examine the effect of the optimization parameter $\epsilon$ on the system performance.
We plot the spectral efficiency of the HM user versus $\epsilon$ in Fig. \ref{epsilon}. The performance of the HM user detection decreases with $\epsilon$ since the function of this parameter is to increase the probability of more LM users receiving data signals from the PBS. Furthermore, the two cases of IDI and full detection converge to similar results as $\epsilon$ increases. This result can provide more clarity on the choice of $\epsilon$ depending on the system requirements and available channel parameters.

{Although the proposed framework provides useful insights into the impact of fractional Doppler on OTFS-NOMA HetNets, several assumptions have been adopted for tractability. In particular, LM users are assumed to perform perfect SIC, while the detection at HM users relies on linear MMSE equalization. Furthermore, although MMSE detection offers a suitable performance and easy implementation, its computational complexity is significant for large OTFS frame dimensions.}

\section{Conclusion}
{In this work, we investigated the OTFS-NOMA paradigm in HetNets consisting of an MBS coordinating multiple PBSs serving mixed-mobility users.} We note that fractional Doppler generates additional IDI, significantly reducing the detection performance of HM users. Therefore, it is essential to account for fractional Doppler when evaluating the performance of OTFS systems. 
On the other hand, the HM user's power allocation factor has been shown to increase the performance gap between the cases with and without IDI. 
Furthermore, the proposed optimization scheme of NOMA power allocation factors improves the fairness between users and enables a higher number of LM users to receive data signals from the PBS at the expense of reduced HM SNR. {The obtained results confirm the effectiveness of OTFS-NOMA in such HetNet environments where multiple HM and LM users coexist. The results showed that neglecting IDI can lead to a spectral efficiency overestimation of up to 0.8 b/s/Hz at a transmit SNR of 25 dB, highlighting the importance of accounting for fractional Doppler effects in OTFS-NOMA systems. The results further showed that full detection can recover most of the performance loss caused by IDI, if the channel parameters in terms of fractional Doppler taps and number of subpaths are known at the receiver.}
{Future research could investigate efficient low-complexity detection techniques to further enhance the performance of OTFS-NOMA systems, particularly under fractional Doppler conditions in high-mobility scenarios.}
\begin{appendices}
\section{Reformulation of the Equalized Signal} 
\label{Gyreformulated}
Since $\mathbf{D}_{a,0} = \mathbf{D}_{{\mathrm{M}},a} +  \mathbf{D}_{{\mathrm{I}},a}$, the matrix $\mathbf{T_0}$ can be reformulated as,
\begin{align}
    \mathbf{T_0} &= \mathbf{\Psi}^{\mathrm{H}} \mathbf{\Delta}_0 \sum_{a = 1}^A v_a \mathbf{D}_{\mathrm{M},a} \mathbf{\Psi} + \mathbf{\Psi}^{\mathrm{H}} \mathbf{\Delta}_0 \sum_{a = 1}^A v_a \mathbf{D}_{\mathrm{I},a} \mathbf{\Psi} \nonumber \\
    &= \mathbf{\Psi}^{\mathrm{H}} \mathbf{\Delta}_0 \mathbf{D}_{\mathrm{M}} \mathbf{\Psi} + \mathbf{F_0} 
     = \mathbf{E_0} + \mathbf{F_0}.
\end{align}
Taking into account that $\mathbf{s} =  \sum_{i = 0}^{U} \sqrt{\beta_i} \mathbf{s}_i$, we get the expression in \eqref{Gy1}.

\section{SNR Expression} \label{snr00}
     The equalized signal can be rewritten as,
\begin{equation}
    \mathbf{G}_0 \mathbf{y}_0 = \mathbf{g}_1 + \mathbf{g}_2 + \mathbf{g}_3 + \mathbf{z}_0,
\end{equation}
where $\mathbf{g}_1 = {\sqrt{\beta_0}} \mathbf{E_0} \mathbf{s}_0  $, $\mathbf{g}_2 =  \mathbf{E_0} \sum_{i = 1}^{U} \sqrt{\beta_i} \mathbf{s}_i $ and $\mathbf{g}_3 =  \mathbf{F}_0 \mathbf{s}$.
The detection SNR of the $k^{\mathrm{th}}$ symbol of the HM user signal is given by,
\begin{equation}
    \gamma_0(k) = \frac{ \mathbb{E} \left( \mathbf{g}_1 \mathbf{g}_1^{\mathrm{H}} \right)_{k,k}}{ \mathbb{E} \left( \mathbf{g}_2 \mathbf{g}_2^{\mathrm{H}} \right)_{k,k} + \mathbb{E} \left( \mathbf{g}_3 \mathbf{g}_3^{\mathrm{H}} \right)_{k,k} + \mathbb{E} \left( \mathbf{z}_0 \mathbf{z}_0^{\mathrm{H}} \right)_{k,k}}.
\end{equation}
Taking into consideration that $\mathbf{s}_u \sim \mathcal{CN}(\mathbf{0}, \beta_u p_q \mathbf{I}_{N \! M})$ and that $\mathbf{w}_u \sim \mathcal{CN}(\mathbf{0},\sigma_\mathrm{w}^2 \mathbf{I}_{N \! M})$, the covariance matrix in the SNR expression can be expressed as,
\begin{align}
    \mathbb{E} \left( \mathbf{g}_1 \mathbf{g}_1^{\mathrm{H}} \right) & = \beta_0 \mathbb{E} \left( \mathbf{E}_0 \mathbf{s}_0 \mathbf{s}_0^{\mathrm{H}}  \mathbf{E}_0^{\mathrm{H}}\right) = \beta_0 P_{\mathrm{T}} \mathbf{\Psi}^{\mathrm{H}} \mathbf{\Delta_E} \mathbf{\Delta_E}^{\mathrm{H}} \mathbf{\Psi} \nonumber \\
    \mathbb{E} \left( \mathbf{g}_2 \mathbf{g}_2^{\mathrm{H}} \right) &= \mathbb{E} \left(\mathbf{E}_0 \left(  \sum_{i = 1}^{U} \sqrt{\beta_i} \mathbf{s}_i \right) \left( \sum_{i = 1}^{U} \sqrt{\beta_i} \mathbf{s}_i^{\mathrm{H}}    \right) \mathbf{E}_0^{\mathrm{H}} \right) \nonumber \\
    &= \sum_{i = 1}^{U} {\beta_i} P_{\mathrm{T}} \mathbf{\Psi}^{\mathrm{H}} \mathbf{\Delta_E} \mathbf{\Delta_E}^{\mathrm{H}} \mathbf{\Psi}  \nonumber \\
    \mathbb{E} \left( \mathbf{g}_3 \mathbf{g}_3^{\mathrm{H}} \right) &= \mathbb{E} \left( \mathbf{F}_0 \mathbf{s} \mathbf{s}^{\mathrm{H}}  \mathbf{F}_0^{\mathrm{H}}\right) =  P_{\mathrm{T}} \mathbf{\Psi}^{\mathrm{H}} \mathbf{\Delta_F} \mathbf{\Delta_F}^{\mathrm{H}} \mathbf{\Psi} \nonumber \\
    \mathbb{E} \left( \mathbf{z}_0 \mathbf{z}_0^{\mathrm{H}} \right) &= \mathbb{E} \left( \mathbf{G}_0 \mathbf{w}_0 \mathbf{w}_0^{\mathrm{H}}  \mathbf{G}_0^{\mathrm{H}}\right) = \sigma_\mathrm{w}^2  \mathbf{\Psi}^{\mathrm{H}} \mathbf{\Delta}_0 \mathbf{\Delta}_0^{\mathrm{H}} \mathbf{\Psi} 
\end{align}
The matrices  $\mathbf{A_1} = \mathbf{\Psi}^{\mathrm{H}} \mathbf{\Delta_E} \mathbf{\Delta_E}^{\mathrm{H}} \mathbf{\Psi}$, $\mathbf{A_2} = \mathbf{\Psi}^{\mathrm{H}} \mathbf{\Delta_F} \mathbf{\Delta_F}^{\mathrm{H}} \mathbf{\Psi}$ and $\mathbf{A_3} = \mathbf{\Psi}^{\mathrm{H}} \mathbf{\Delta}_0 \mathbf{\Delta}_0^{\mathrm{H}} \mathbf{\Psi}$ are block circulant matrices. Therefore, their diagonal elements are equal, i.e, $\left(\mathbf{A_1} \right)_{k,k} = a_1$, $\left(\mathbf{A_2} \right)_{k,k} = a_2$ and $\left(\mathbf{A_3} \right)_{k,k} = a_3$ for all $1 \le k \le NM$.
Moreover, since $\mathbf{\Psi}$ is unitary, $\mathrm{tr}(\mathbf{A_1}) = \mathrm{tr}(\mathbf{\Delta_E} \mathbf{\Delta_E}^{\mathrm{H}})$, $\mathrm{tr}(\mathbf{A_2}) = \mathrm{tr}(\mathbf{\Delta_F} \mathbf{\Delta_F}^{\mathrm{H}})$ and $\mathrm{tr}(\mathbf{A_3}) = \mathrm{tr}(\mathbf{\Delta}_0 \mathbf{\Delta}_0^{\mathrm{H}})$. Thus, $\sum_{k = 1}^{NM} \left(\mathbf{A_1} \right)_{k,k} = NM a_1 = \sum_{k = 1}^{NM} \left| \delta_{\mathrm{E},k}\right|^2$, where $\delta_{\mathrm{E},k}$ are the elements of $\mathbf{\Delta_E}$. As a result, $a_1 = \sum_{k = 1}^{NM} \frac{ \left| \delta_{\mathrm{E},k}\right|^2 }{NM}$. We apply the same derivations to $\mathbf{A_2}$ and $\mathbf{A_3}$.
We normalize the SNR by the noise power and deduce the final expression in \eqref{snr0}.

\section{LASSO Optimization} 
\label{optimization}
 The optimal solutions satisfy the following conditions,
\begin{equation}\label{delta}
    \frac{\delta L}{\delta \beta_u} = 0, \forall u=0, \ldots, U.
\end{equation}
\begin{align}
\begin{cases}
    \frac{\delta L}{\delta \beta_0} &= -1 + \theta_2 = 0 \qquad \quad  \; \; \qquad\textrm{\textbf{Result 1}} \nonumber \\
    \frac{\delta L}{\delta \beta_u} &= - \frac{\lambda_\mathrm{L} + \theta_1}{\log(2)} \frac{a_u}{1+a_u \beta_u} + 1 = 0 \quad \textrm{\textbf{Result 2}}
    \end{cases}
\end{align}
The results of equation \eqref{delta} are the following:
\begin{itemize}
    \item \textbf{Result 1}: we derive $\theta_2 = 1$, therefore, constraint C2 is active, i.e. $\sum_{u = 0}^{U} \beta_u = 1$.
    \item \textbf{Result 2}: $1 + a_u \beta_u = \frac{\lambda_\mathrm{L} + \theta_1}{\log(2)} a_u$. 
\end{itemize}
From \textbf{Result 1}, $\beta_0 = 1 - \sum_{u = 1}^{U} \beta_u $.\\
From \textbf{Result 2}, there are two cases for $\theta_1$:
\begin{itemize}
    \item If $\theta_1 \neq 0$, then constraint C1 is active. Thus, $\theta_1 = X_U \log(2) - \lambda_{\mathrm{L}}$ and $\beta_u = X_U  - 1/a_u, 1 \le u \le U$, where $X_U$ is expressed in \eqref{XU}. Since $\theta_1 > 0$, then $X_U \log(2) > \lambda_{\mathrm{L}}$.
    \item If $\theta_1 = 0$, then constraint C1 is inactive. Thus, $\sum_{u = 1}^{U} \log_2 (1 + a_u \beta_u) > R_{\min}$ equivalent to $X_U \log(2) < \lambda_{\mathrm{L}}$. Moreover, $\beta_u = \frac{\lambda_{\mathrm{L}}}{\log(2)} - 1/a_u, 1 \le u \le U$.
\end{itemize}
As a result, the optimal solution for $1 \le u \le U$ can be expressed as, 
\begin{align}
    \beta_u = 
    \begin{cases}
        & X_{U} - {1}/{a_u}, \quad \mathrm{if} \; \lambda_{\mathrm{L}} \le X_{U} \log(2), \\
        & \frac{\lambda_{\mathrm{L}}}{\log(2)}- {1}/{a_u}, \quad \mathrm{otherwise}.
    \end{cases}
\end{align}

\end{appendices}

\bibliographystyle{IEEEtran}
\bibliography{IEEEabrv,main}

\end{document}